\input epsf.tex

\documentclass[preprint,pra,aps,showpacs,preprintnumbers,amsmath,amssymb,eqsecnum]{revtex4}

\def\la{{\langle}}
\def\ra{{\rangle}}
\def\a{{\alpha}}

\def\l{{\lambda}}

\def\x{{\mathbf x}}
\def\y{\mathbf{y}}

\def\p{\mathbf{p}}

\def\a{\alpha}

\def\half{\frac{1}{2}}

\newcommand\beq{\begin{equation}}
\newcommand\eeq{\end{equation}}
\newcommand\bea{\begin{eqnarray}}
\newcommand\eea {\end{eqnarray}}

\begin{document}

%\preprint{APS/123-QED}

\title{Probabilities in Quantum Cosmological Models:\\ A Decoherent Histories
Analysis Using a Complex Potential}

\author{J.J.Halliwell}%


\affiliation{Blackett Laboratory \\ Imperial College \\ London SW7
2BZ \\ UK }
%\affiliation{

\date{\today}% It is always \today, today,

\begin{abstract}
In the quantization of simple cosmological models (minisuperspace
models) described by the Wheeler-DeWitt equation, an important
step is the construction, from the wave function, of a probability
distribution answering various questions of physical interest,
such as the probability of the system entering a given region of
configuration space at any stage in its entire history.
A standard but heuristic procedure is to use
the flux of (components of) the wave function in a WKB
approximation. This gives sensible semiclassical results but lacks
an underlying operator formalism. In this paper, we address the
issue of constructing probability distributions linked to the
Wheeler-DeWitt equation using the decoherent histories approach to
quantum theory. The key step is the construction of class
operators characterizing questions of physical interest. Taking
advantage of a recent decoherent histories analysis of the arrival
time problem in non-relativistic quantum mechanics, we show that
the appropriate class operators in quantum cosmology are readily
constructed using a complex potential. The class operator
for not entering a region of configuration space is given
by the $S$-matrix for scattering off a complex potential localized
in that region. We thus derive the class operators for entering
one or more regions in configuration space. The class operators
commute with the Hamiltonian, have a sensible
classical limit and are closely related to an intersection
number operator.
The definitions of class operators
given here handles the key case in which the underlying classical
system has multiple crossings of the boundaries of the regions
of interest.
We show that oscillatory WKB solutions to
the Wheeler-DeWitt equation give approximate decoherence of histories,
as do superpositions of WKB solutions, as long as the regions
of configuration space are sufficiently large.
The
corresponding probabilities coincide, in a semiclassical approximation,
with standard heuristic
procedures. In brief, we exhibit the well-defined operator formalism
underlying the usual heuristic interpretational methods in quantum cosmology.

\end{abstract}

\maketitle

\section{Introduction}

\subsection{Opening Remarks}

A question in quantum gravity that continues to attract considerable interest is the
problem of quantizing and interpreting the Wheeler-DeWitt equation of minisuperspace
quantum cosmology \cite{QC1,QC,QC2,DeW,Rov},
\beq
H \Psi = 0
\label{1.1}
\eeq
Although the setting of this problem is simple cosmological models with just
a handful of homogeneous parameters, the techniques employed in answering this question
may be relevant to general approaches to quantum gravity, such as the loop variables approach
or causal set approach. This is because the central difficulty in consistently
quantizing and interpreting the Wheeler-DeWitt equation is the absence of a variable to play
the role of time and all approaches to quantum gravity must confront this issue at some stage
\cite{Time}.

A frequently studied example consists of a closed FRW cosmology with scale factor $a=e^{\a}$ and
a homogeneous scalar field $\phi$ with (inflationary) potential $V( \phi)$ \cite{FRW}.
The Wheeler-DeWitt equation
for this model is
\beq
\left( \frac { \partial^2} { \partial \a^2} - \frac { \partial^2} {\partial \phi^2 }
+ e^{ 6 \a } V( \phi) - e^{4 \a} \right) \Psi ( \a, \phi) = 0
\label{1.2}
\eeq
Given suitable boundary conditions, one can solve this equation for the wave function $\Psi (\a, \phi)$
and attempt to use it to answer a number of interesting cosmological questions. There
are many such questions: Is there a regime in which the universe behaves approximately
classically?
What is the probability that the universe expands beyond a given size $a_0$?
What is the probability that the universe has a certain energy density at a given value
of the scale factor? What is the probability that the universe's history
passes through a given region $\Delta$ of configuration space, characterized by certain ranges of $a$ and $\phi$?

Such questions, of necessity, do not involve the specification of an external
time. Classically, the system's trajectories in minisuperspace
may be written as paths $ (\alpha (t), \phi (t) )$, but here $t$ is a convenient but
unphysical parameter that labels the points along the paths -- it does not correspond to the physical
time measured by an external clock. The absence of a physical time is reflected in the
quantum theory by the fact that the quantum state obeys the Wheeler-DeWitt equation
Eq.(\ref{1.1}), not a Schr\"odinger equation, and it is this difference that presents
such a challenge to conventional quantization methods.

Although very plausible heuristic semiclassical methods exist for formulating and answering the above
questions (in particular, the WKB interpretation \cite{QC1}),
it is of interest to see whether or not there exists a precise and well-defined quantum-mechanical
scheme underlying these heuristic methods. We are not looking for high standards of
mathematical rigour -- just a standard quantum-mechanical framework of operators, inner product
structures etc, obeying some reasonable requirements.
The purpose of this paper is to show, building on earlier attempts \cite{Har3,HaMa,HaTh1,HaTh2,HaWa,CrHa,
Whe,AnSa},
that such a framework exists in the context of the decoherent
histories approach to quantum theory.

\subsection{Inner Products and Operators for the Wheeler-DeWitt Equation}

In the Wheeler-DeWitt equation (\ref{1.1}), $H$ is the Hamiltonian operator of a minisuperspace model with $n$ coordinates
$q^a$, and is typically of the form
\beq
H = -  \nabla^2 + U (q)
\label{1.3}
\eeq
where $ \nabla $ is the Laplacian in the minisuperspace metric $f_{ab}$, which has signature $ ( - + + + \cdots )$.
It is naturally linked to the current,
\beq
J_{a} =  i \left( \Psi^* \overrightarrow \partial_{a} \Psi - \Psi^* \overleftarrow \partial_a \Psi \right)
\eeq
which is conserved
\beq
\nabla \cdot J = 0
\eeq
Closely associated is the Klein-Gordon inner product defined on a surface $\Sigma$
\beq
( \Psi, \Phi)_{KG} =  i \int_{\Sigma} d \sigma^a
\left( \Psi^* \overrightarrow \partial_{a} \Phi - \Psi^* \overleftarrow \partial_a \Phi \right)
\label{1.6}
\eeq
where $ d \sigma^a$ is a surface element normal to $\Sigma$. In flat space with a constant potential,
the Wheeler-DeWitt equation is just the Klein-Gordon equation. Its solutions may be sorted out
into positive and negative frequency in the usual way. With a little attention to sign, it
is then possible to use components of the current to define probabilities.

However, in general, it is not possible to sort the solutions to the Wheeler-DeWitt equation
into positive and negative frequency. This is one manifestation of the problem of time
and more elaborate methods are required to associate probabilities with the Wheeler-DeWitt equation.
There are two main issues: finding an inner product, and then finding suitable operators.

The issue of finding a suitable positive inner product is reasonably straightforward
and goes by the name of Rieffel induction, or the induced (or physical) inner product \cite{HaMa,Rie}.
We consider first the usual Schr\"odinger inner product,
\beq
\la \Psi_1 | \Psi_2 \ra = \int d^n q \ \Psi_1^* (q) \Psi_2 (q )
\eeq
We then consider eigenvalues of the Wheeler-DeWitt operator
\beq
H | \Psi_{\l k} \ra = \l | \Psi_{\l k} \ra
\eeq
where $k $ labels the degeneracy. These eigenstates
will satisfy
\beq
\la \Psi_{\l' k'} | \Psi_{\l k} \ra =
\delta (\l - \l') \delta (k - k')
\eeq
from which it is clear that this inner product diverges
when $\l = \l'$. The induced inner product on a set of
eigenstates of fixed $\l$ is defined, loosely speaking,
by discarding the $\delta$-function $\delta (\l-\l')$.
That is, the induced (or physical) inner product is
then defined by
\beq
\la \Psi_{\l k'} | \Psi_{\l k} \ra_{phys} =
\delta (k-k')
\eeq
This procedure can be defined quite rigorously, and has been
discussed at some length in Refs.\cite{Rie,HaMa}. It is readily
shown that the induced inner product coincides with the Klein-Gordon
inner product when a division into positive and negative frequencies
is possible, with the signs adjusted to make it positive. (This is described
in Appendix A).

Turning now to the construction of interesting operators,
the interesting dynamical variables associated with the Wheeler-DeWitt
equation are those that commute with $H$. This is because the constraint equation
is related to reparametrization invariance -- which is reflected in the absence
of a physical time variable -- and we seek operators which are invariant.
A wide class of operators commuting with $H$ are of the form
\beq
A = \int_{-\infty}^{\infty} dt  \ B ( t )
\label{1.11}
\eeq
which clearly commute with $H$ since
\bea
e^{i H s } A e^{ - i H s } &=& \int_{-\infty}^{\infty} dt \ B ( t + s)
\nonumber \\
&=& A
\eea
Many examples are given in Refs.\cite{Rov,Rie,HaTh2}. However, another way of constructing
such operators involves taking products,
\beq
A = \prod_{t= - \infty}^{\infty} B ( t)
\eeq
which may be shown to commute with $H$ using essentially the same argument \cite{HaWa},
but clearly further mathematical detail is required to give meaning to the infinite
product. (Here $t$ is the unphysical parameter time).

Given these prescriptions for inner products and operators, one may then attempt
to construct operators and probabilities implementing some of the questions mentioned above.
We will focus on the following general question: Given a solution $\Psi $ to the Wheeler-DeWitt
equation, what
is the probability of finding the system in a
region $\Delta$ of configuration space, or of crossing a surface $\Sigma$,
at any stage in the system's history? The question is depicted in Figure 1.

Questions involving surface crossings are not unlike more familiar questions in
non-relativistic quantum mechanics,
where there is a physical time parameter, but the key difference in quantum cosmology is that even classically,
a given trajectory will typically cross a fixed surface more than once.
It is precisely these sorts of issues that
need to be carefully phrased in the quantum theory. Whilst the operator
formalism briefly outlined above has been used to address such questions \cite{Rov}, the problems of
characterizing properties of trajectories and surface crossings in quantum theory is naturally
accommodated in the decoherent histories approach to quantum theory.

\subsection{The Decoherent Histories Approach}

A more general approach to quantizing and interpreting the Wheeler-DeWitt equation
is provided by the decoherent histories approach to quantum theory, and this
approach will be the focus of this paper \cite{GH1,GH2,Gri,Omn1,Omn2,Hal1,Hal2,Halrec,Ish}. In this approach, probabilities
are assigned to histories using the formula
\beq
p(\a) = {\rm Tr} \left( C_{\a} \rho C_{\a}^\dag \right)
\label{1.15}
\eeq
Here, $\rho$ is the initial state (in our case a pure state) and
$C_{\a}$ is a class operator characterizing the histories $\a$ of interest.
In non-relativistic quantum mechanics these class operators are given
by time-ordered strings of projection operators,
\beq
C_{\a} = P_{\a_n} (t_n) \cdots P_{\a_2} (t_2) P_{\a_1} (t_1)
\label{1.16}
\eeq
(or by sums of such strings). The class operators alway satisfies the condition
\beq
\sum_{\a} C_{\a} = 1
\label{1.17}
\eeq
For the reparamterization invariant
systems considered here, the definition of the class operators is more subtle,
and we return to this below.

Because of interference between pairs of histories, probabilities cannot always
be assigned. To check for interference we therefore consider the decoherence
functional,
\beq
D(\a, \a') =  {\rm Tr} \left( C_{\a} \rho C_{\a'}^\dag \right)
\eeq
When
\beq
D(\a, \a') = 0
\eeq
for all pairs of histories in the set with $\a \ne \a'$, we say that there is decoherence of the set of histories and probabilities may be assigned Eq.(\ref{1.15}).
When there is decoherence, it is easily seen from Eq.(\ref{1.17}) that the probabilities
Eq.(\ref{1.15}) are also given by
\beq
p(\a) = {\rm Tr} \left( C_{\a} \rho \right) = {\rm Tr} \left( C_{\a}^{\dag} \rho
\right)
\label{1.20}
\eeq
Decoherence guarantees that this expression is real and positive, even though
the class operators are not positive or hermitian operators in general.
% Comments on this

The structure of the decoherent histories approach is very general and may be applied
to a wide variety of situations, given an initial state, class operators, and a suitable
inner product structure. For the application to the Wheeler-DeWitt system considered
here, the initial state is take to be a solution to the Wheeler-DeWitt equation
and the inner product is the induced inner product described above.

The most crucial
element is the specification of the class operators $C_{\a}$. In non-relativistic
quatum theory, the class operator
is generally a string of projection operators like Eq.(\ref{1.16}),
or a sum of strings of such operators, but this basic structure can be
generalized in various ways. The product in the string can be
taken to be continuous time \cite{Ish}. Also, it is often valuable, sometimes essential,
to allow the projectors to be replaced by more general operators, such as POVMs.
The class operators must also properly characterize the histories one is interested
in. It is not always obvious how to do this but
useful clues often come from looking at the classical analogue
of the class operator (where all the projectors commute).

Here, we are interested in histories which enter a region $\Delta$ of configuration space,
or which cross a surface $\Sigma$, but without regard to time. This absence
of a physical time variable seems particularly challenging, given that time seems
to be central to the definition of non-relativistic analogue, Eq.(\ref{1.16}).
Closely related to this is the role of the constraint equation, Eq.(\ref{1.1}).
As noted already,
these two features are directly related to the underlying symmetry of the theory
-- reparametrization invariance -- and this symmetry is
respected if the class operators commute with the constraint,
\beq
[H, C_{\a} ] = 0
\label{1.20a}
\eeq
Eq.(\ref{1.20a}) is in keeping with standard procedures of Dirac quantization \cite{Rov})
and also ensures that the class operators have a sensible classical limit \cite{HaWa}.

Some partially successful
attempts to construct such class operators have been given previously \cite{HaMa,HaTh1,HaTh2,HaWa}, but
ran into various problems to do with the Zeno effect and with compatibility
with the constraint equation.
The main
aim of this paper is to give fully satisfactory definitions of class operators
for quantum cosmological models and explore their decoherence properties and
probabilities.

\subsection{Some Properties of the Wheeler-DeWitt Equation and the WKB Interpretation}

To prepare the way for the full decoherent histories analysis of quantum cosmology,
it is important to discuss some properties of the Wheeler-DeWitt equation and
review the commonly used heuristic semiclassical interpretation
of the wave function, since a proper quantization must recover this structure
in some limit.

The Wheeler-DeWitt equation Eq.(\ref{1.3}) is a Klein-Gordon equation in
a curved configuration space with indefinite metric $f_{ab} (q)$ and
potential $U(q)$ which can be positive or negative. The curvature effects of the metric are not
significant in relation to the issues addressed in this paper, so we will assume
for simplicity that the metric is flat.

The classical constraint equation corresponding to the Wheeler-DeWitt equation
Eq.(\ref{1.3}) is
\beq
\frac {1} {4} f_{ab} \dot q^a \dot q^b + U = 0
\eeq
from which one can see that the classical trajectories are timelike in the region
$ U>0$ and spacelike in $U<0$. (The timelike direction is that of increasing
scale factor and the spacelike directions correspond to matter and anisotropic modes).
The quantum case has analogous features.
In simple models such as Eq.(\ref{1.2}), the
character of the solutions to the Wheeler-DeWitt equation depends on the sign of
$U$. For large scale factors, $U>0$ and the wave function is oscillatory,
corresponding, very loosely, to a quasiclassical regime, and for small scale factors, $U<0$,
and the wave function is exponential, corresponding to a classical forbidden regime.
However, there are certain types of models (such as those with an exponential
potential for the scalar field), in which the identification of the oscillatory
and exponential regions depends also on whether the constant $U$ surfaces are spacelike
or timelike \cite{HalKK}. We will not address this here.

One can also associate a Feynman propagator $G_F$ with the Wheeler-DeWitt operator,
\bea
G_F &=& \int_0^{\infty} dt \ e^{  - i H t - \epsilon t}
\nonumber \\
&=&\frac { - i } { H  - i \epsilon }
\label{GF}
\eea
where $\epsilon \rightarrow 0+$. (Numerous propagator-like objects of this type have been considered
in quantum cosmology \cite{Teit}).
Locally, on scales smaller than the scale on which the potential
$U$ significantly varies, the propagator $G_F (q,q')$ will be
essentially identical in its properties with the analogous object for the relativistic
particle for flat space. Therefore, in the region $U>0$, for points
$q$ and $q'$ which are timelike separated, it will propagate
positive frequency solutions to the future and negative frequency solutions to the
past. It will be exponentially suppressed for initial and final points $q$ and $q'$
which are spacelike separated (this is the familiar ``propagation outside the lightcone''
effect). For $U<0$, similar statements hold but with timelike
and spacelike reversed. Similar features hold on larger scales in a semiclassical
approximation.
These properties are important to understand the class operator
constructed below.

Very plausible but heuristic answers to questions concerning crossing surfaces and
entering regions
are readily found using the
WKB approximate solutions to the Wheeler-DeWitt equation and the Klein-Gordon current
\cite{QC1}.
In the oscillatory regime, the Wheeler-DeWitt equation may be solved using the
WKB ansatz
\beq
\Psi = R e^{ i S}
\label{WKB}
\eeq
where the rapidly varying phase $S$ obeys the Hamilton-Jacobi equation
\beq
( \nabla S)^2 + U = 0
\label{1.24}
\eeq
and the slowly varying prefactor $R$ obeys
\beq
\nabla \cdot ( | R |^2 \nabla S ) = 0
\label{1.26}
\eeq
The latter equation is just current conservation for the WKB current,
\beq
J = | R | ^2 \nabla S
\eeq
Wave functions of the WKB form Eq.(\ref{WKB}) indicate a correlation between
position and momenta of the form
\beq
p = \nabla S
\label{ps}
\eeq
and this suggests that the wave function Eq.(\ref{WKB}) corresponds to an ensemble of classical trajectories
satisfying Eq.(\ref{ps}). The current $J$ may then be used to define a measure on this
set of trajectories.
For example, we consider a surface $\Sigma$ and choose the normal $n^a$ to the surface
so that $ n \cdot \nabla S > 0 $. Then the probability of the system crossing
the surface is taken to be
\beq
p (\Sigma ) = \int_{\Sigma} d^{n-1} q \ n^a J_a
\eeq

Of particular interest here is the probability of entering a region $\Delta$. This is clearly related
to the flux at the boundary of the region. The current will typically
intersect the boundary of $\Delta$ {\it twice}. However, we can split the boundary $\Sigma$
of $\Delta$ into two sections: $\Sigma_{in}$ at which the current is ingoing and $\Sigma_{out}$
at which the current is outgoing. The probability of entering $\Delta$ is may then be
expressed in the two forms
\bea
p (\Delta ) &=& - \int_{\Sigma_{in}} d^{n-1} q \ n^a J_a
\nonumber \\
 &=&  \int_{\Sigma_{out}} d^{n-1} q \ n^a J_a
\label{1.29}
\eea
where here we have defined the normal $n^a$ to point outwards. See Figure 2.
These two forms are equivalent since the current is locally conserved.

Typically, little is said of the regions in which the wave function is exponential, except that
they are similar to tunneling regions in non-relativistic quantum mechanics, so are
classically forbidden in some sense. However, the wave function is not necessarily
small in these regions, so there is surely more to it than this. In Ref.\cite{Halcor}, it was noted
that, unlike the oscillatory regions, the exponential regions do not indicate
a correlation between position and momenta of the form Eq.(\ref{ps}), and it seems
that this should be significant in some way.

Whilst the WKB interpretation is very plausible and adequate for most situations
of interest, it leaves many questions unanswered. The main issue is to understand
the operator origin of these probabilities, in terms of the language of operators commuting
with $H$ developed above. Furthermore, what can one say about superpositions of WKB states?
Can interference between them be neglected? Also, what can one say about the exponential,
classically forbidden regions? Is there a more precise way of saying that they
are non-classical?

%The important of a formalism that handles multiple intersections.
%Relation to the intersection number operator.

\subsection{This Paper}

The purpose of this paper is to present a decoherent histories quantization
of the Wheeler-DeWitt equation, and in particular to
exhibit exactly defined class operators which characterize
histories entering regions of configuration space or crossing surfaces,
without reference to an external time. The key idea is to use
a complex potential to define the class operator for not entering
a region $\Delta$ in configuration space. In particular, we take the class operator
for not entering to be the $S$ matrix describing scattering off a complex
potential localized in $\Delta$. This turns out to have all the right properties
and to give physically sensible results for decoherence and probabilities.

In Section 2, to explain and motivate the use of a complex potential we review the use
of such potentials in the decoherent histories analysis of the arrival
time problem in non-relativistic quantum mechanics. In Section 3, we
describe the construction of class operators for the Wheeler-DeWitt equation
using a complex potential. The properties of the class operator
for entering a region $\Delta$ are described in Section 4. The class operator
has a sensible classical limit which, crucially, registers just one intersection
of a trajectory with a surface, even when the trajectory intersects twice or more.
In the quantum case, the class operator is an operator describing the
ingoing flux across the boundary of the region, an expected
result on semiclassical grounds, and is closely related to an intersection
number operator.

In Section 5, some simple one and two dimensional examples are considered in detail
to confirm the properties of the class operator outlined in Section 4. We also confirm
that the formalism gives sensible and expected results for surface crossings
of the relativistic particle
in flat space. In Section
6 we consider the WKB regime and show that the decoherent histories analysis
reproduces the expected heuristic interpretation of the wave function.
We summarize and conclude in Section 7. Some properties of the inner product
structure of the Klein-Gordon equation are summarized in Appendix A.

%[Does not involve a global definition of time. Just a local definition of surface
%crossing.]

\section{The Arrival Time Problem in Non-Relativistic Quantum Theory}

In this section we summarize some of the key features of
the arrival time problem in non-relativistic quantum mechanics \cite{timeinqm}.
These details are very relevant to the quantum cosmology case and
in particular motivate the use of complex
potentials in the definition of class operators.

In the one-dimensional statement of the arrival time problem, one considers an
initial wave function $ | \psi \rangle $ concentrated in the
region $x>0$ and consisting entirely of negative momenta. The
question is then to find the probability $ \Pi (\tau) d \tau $
that the particle crosses $x=0$ between time $ \tau$ and $ \tau +
d \tau$. See Figure 3. The canonical answer is the current density
\beq
J(\tau) = \frac
{(- 1) } {2 m } \langle \psi_\tau | \left( \hat p \delta ( \hat x
) + \delta ( \hat x ) \hat p \right) | \psi_\tau \rangle
\label{2.1}
\eeq
where $ | \psi_\tau \rangle $ is the freely
evolved state.
This has the correct semiclassical limit, but can be negative for
certain types of states consisting of superpositions of different
momenta (backflow states).
It is of interest to explore whether this simple semiclassical result
emerges from more elaborate measurement-based
or axiomatic schemes. Many such schemes naturally lead to the use of a complex potential
$ - i V (x) $ in the Schr\"odinger equation \cite{complex}, where
\beq
V(x) =
V_0 \theta (-x)
\label{2.2}
\eeq
With such a potential, the state
at time $\tau$ is
\beq
| \psi (\tau) \rangle = \exp \left( - i H
\tau -  V_0  \theta (-x) \tau \right) | \psi \rangle
\label{2.3}
\eeq
where
$H$ is the free Hamiltonian. The idea here is
that the part of the wave packet that reaches the origin during
the time interval $[0, \tau]$ should be absorbed, so that
$ \langle \psi (\tau) | \psi (\tau) \rangle $
is the
probability of not crossing $x=0$ during the time interval $[0,\tau]$. The
probability of crossing between $ \tau$ and $\tau + d \tau $ is
then
\beq
\Pi (\tau) = - \frac { d  } { d \tau} \langle \psi (\tau) | \psi (\tau) \rangle
\label{2.5}
\eeq
This may be approximately evaluated in the limit $V_0 \ll E $ (where $E$ is the typical energy scale),
which is the limit of negligible reflection off the complex potential.
The result is
\beq
\Pi ( \tau) =
 2 V_0 \int_{0}^\tau dt \ e^{ - 2V_0 (\tau - t) } \ J(t)
\eeq
The probability for crossing during a finite interval $[\tau_1, \tau_2]$ is then given by
\beq
p (\tau_2, \tau_1) = \int_{\tau_1}^{\tau_2} d \tau \ \Pi (\tau)
\eeq
and if this time interval is sufficiently large compared to $ 1 / V_0$, we have
\beq
p (\tau_2, \tau_1) \approx  \int_{\tau_1}^{\tau_2} dt \ J(t)
\label{2.8}
\eeq
which means that the dependence on the potential drops out entirely at sufficiently
coarse grained scales, and there is agreement with semiclassical expectations
involving the current $J(t)$.

In the decoherent histories analysis of the arrival time problem \cite{HaYe1,HaYe2}, we first
consider the construction of the class operator $C_{nc}$ for not crossing
the origin during the finite time interval $[0, \tau]$.
We split the time interval into $N$ parts
of size $\epsilon$, and the class operator is provisionally defined by
\beq
C_{nc} = P e^{ - i H \epsilon } P \cdots e^{ - i H \epsilon } P
\label{2.9}
\eeq
where there are $N+1$ projections $P = \theta ( \hat x)$ onto the positive
$x$-axis and $N$ unitary evolution operators in between. One might be tempted
to take the limit $ N \rightarrow \infty $ and $ \epsilon \rightarrow 0 $,
but this yields physically unreasonable
results. This limit actually yields the restricted propagator in $x>0$,
\bea
C_{nc} &=& g_r (\tau, 0 )
\nonumber \\
&=& P e^{ - i P H P \tau }
\label{2.9b}
\eea
This object is also given by the path integral expression
\beq
\langle x_1 | g_r (\tau, 0 ) | x_0 \rangle = \int_r{\cal D} x  \exp ( i S )
\label{rest}
\eeq
where the integral is over all paths from $ x(0) = x_0$ to $x(\tau) = x_1$
that always remain in $x(t) > 0 $. However, the class operator
defined by Eq.(\ref{2.9b}) has a problem with
the Zeno effect -- it consists of continual projections onto the region
$x>0$ and as a result the wave function never leaves the region. This is reflected
in the fact that the restricted propagator $g_r$ is unitary in the Hilbert
space of states with support only in $x>0$. This is a serious difficulty which has
plagued a number of earlier works in this area \cite{YaT,HaZa,Har4}.

To avoid the Zeno effect, the key is to keep $\epsilon $ finite, and in particular,
standard wisdom suggests that $ \epsilon > {1} / { \Delta H } $,
where $H$ is the free particle Hamiltonian. The class operator Eq.(\ref{2.9})
is not easy to work with for finite $ \epsilon $, but fortunately
an extremely useful result of Echanobe et al. comes to the rescue \cite{Ech}.
They argued that the string of operators in Eq.(\ref{2.9}) is in fact approximately
equivalent to evolution in the presence of the complex potential $ - i V$ introduced
above.
That is,
\beq
P e^{ - i H \epsilon } P \cdots e^{ - i H \epsilon } P
\ \approx \ \exp \left( - i H_0 \tau - V_0 \theta (- \hat x ) \tau \right)
\label{2.12}
\eeq
They argued that this is valid for $ \Delta H \ll V_0 $ and $V_0 \epsilon  \gg 1$,
but there is evidence that this connection is valid more generally \cite{Ye1}.
In any event it strongly suggests that the class operator $C_{nc}$, normally
defined by a string of projection operators, is justifiably defined instead
using a complex potential. That is, we define
\beq
C_{nc} = \exp \left( - i H_0 \tau - V_0 \theta (- \hat x ) \tau \right)
\label{2.13}
\eeq
The subsequent decoherent histories analysis was described in detail in Ref.\cite{HaYe1}.
The corresponding class operator for crossing during a time interval $[\tau_1, \tau_2]$ was
shown to be
\beq
C_c (\tau_2, \tau_1) = \int_{\tau_1}^{\tau_2} dt \ e^{- i H_0 (\tau - t)} V e^{- i H_0 t - V t}
\label{7.7}
\eeq
and again in the approximation $ V_0  \ll E $ and for time intervals
greater than $1 / V_0$ this may be shown to have the simple
and appealing form
\beq
C_c (\tau_2, \tau_1) \approx  \int_{\tau_1}^{\tau_2} dt \ e^{ - i H \tau} \frac {(-1)} {2m}
\ \left( \hat p \ \delta (\hat x_t)  \    +
\ \delta (\hat x_t ) \ \hat p \right)
\label{7.15}
\eeq
which is now independent of the $V_0$. (These definitions of class operators differ
by a factor of $\exp ( - i H \tau) $ from those defined in Section 1C, but this does
not make any difference in the decoherence functional and probabilities).
With this class operator, one can
show that
there is decoherence of histories for a variety of interesting initial states, such as wave packets
(but not for superposition states with backflow), and for such states, the general result Eq.(\ref{1.20})
implies that the probability for crossing is simply
\beq
p (\tau_2, \tau_1) = \langle \Psi | C_c (\tau_2, \tau_1) | \Psi \rangle
\eeq
which agrees precisely with the semiclassically expected result, Eq.(\ref{2.8}).

In summary, the decoherent histories analysis of the arrival time problem in non-relativistic
quantum mechanics indicates that it is reasonable to define class operators for not entering
a space time region using a complex potential, as in Eq.(\ref{2.13}), and that such a definition
gives sensible a semiclassical limit, independent of the potential,
at sufficiently coarse grained scales.

\section{Construction of the Class Operators Using a Complex Potential}

We now come to the central issue concerning this paper, which is the construction
of class operators for the decoherent histories analysis of the Wheeler-DeWitt equation.

\subsection{Class Operators for a Single Region}

We seek
class operators for the system Eq.(\ref{1.3}) describing histories
which enter or do not enter the region $\Delta$,
without specification of the time at which they enter.
It is easiest to first focus on the class operator $\bar C_{\Delta}$ for not entering
and the class operator for entering is then given by
\beq
C_{\Delta} = 1 - \bar C_{\Delta}
\eeq

The earliest attempts to define class operators for the Wheeler-DeWitt equation
involved defining $\bar C_{\Delta}$ as a sort of propagator obtained by
integrating restricted propagators of the form Eq.(\ref{rest}) over
an infinite range $t$ (now regarded as the unphysical parameter time) \cite{HaMa,HaTh1,HaTh2}.
However, in addition to having problems with the
Zeno effect (which was not in fact appreciated in these earlier works),
such constructions are difficult to reconcile with the constraint equation
and some ad hoc modifications of the basic construction were required
to give sensible answers.

A rather different approach to constructing
$\bar C_{\Delta}$ was given in Ref.\cite{HaWa}. This was again problematic,
but we review the
construction here, since it is readily modified to yield a successful definition
of the class operators.
We denote
by $P$ the projector onto $\Delta$ and $\bar P$ the projector onto
the outside of $\Delta$.
Our provisional proposal for the class operator
for trajectories not entering $\Delta$ is the time-ordered infinite product,
\begin{equation}
\bar C_{\Delta} = \prod_{t=-\infty}^{\infty} \bar P (t)
\label{3.13}
\end{equation}
where $t$ is the unphysical parameter time.
Subject to a more precise definition, given shortly, this object has
the required properties. It is a string of projectors. Classically, it is equal
to one for trajectories which remain outside $\Delta$ at every moment of parameter
time. Also, it commutes with $H$, at least formally.

To define this more precisely, we first consider the product of
projectors at a discrete set of times, $ t_1 $, $t_1 + \epsilon$,
$t_1 + 2 \epsilon,  \cdots, t_1 + n \epsilon = t_2$.
We define the intermediate quantity, $ \bar C_{\Delta} (t_2,t_1)$ as the continuum limit
of the product of projectors,
\begin{equation}
\bar C_{ \Delta} (t_2,t_1) = \lim_{\epsilon \rightarrow 0} \bar P (t_2) \bar P(t_2
-\epsilon) \dots \bar P(t_1 + \epsilon ) \bar P (t_1)
\label{3.15}
\end{equation}
where the limit is $n \rightarrow \infty$, $\epsilon \rightarrow 0 $ with
$ t_2 - t_1$ fixed. The desired class operator is then
\begin{equation}
\bar C_{\Delta} = \lim_{t_2 \rightarrow \infty, t_1 \rightarrow -\infty} \ \bar C_{ \Delta} (t_2,t_1)
\label{3.16}
\end{equation}

The class operator operator is clearly closely related to the restricted propagator $g_r$ (the generalization
of Eq.(\ref{rest}))
in the region outside $\Delta$,
since we have
\begin{equation}
\bar C_{\Delta} (t_2,t_1) = e^{ i H t_2} \ g_r (t_2,t_1) \ e^{ - i H t_1}
\label{3.17}
\end{equation}
and therefore
\begin{equation}
\bar C_{ \Delta} = \lim_{t_2 \rightarrow \infty, t_1 \rightarrow -\infty} \  e^{ i H t_2} \ g_r (t_2,t_1) \ e^{ - i H t_1}
\label{3.18}
\end{equation}
The class operator commutes with $H$. This is because,
from Eq.(\ref{3.17})
\begin{equation}
e^{ i H s} \bar C_{ \Delta} (t_2,t_1) e^{ - i H s} = e^{ i H (t_2+s)}
\ g_r (t_2,t_1) \ e^{ - i H (t_1+s)}
\label{3.18b}
\end{equation}
This becomes independent of $s$ as $ t_2 \rightarrow \infty$, $t_1 \rightarrow - \infty$,
hence
\begin{equation}
[ H, \bar C_{ \Delta} ] = 0
\end{equation}

The problem with this definition, however, is that it suffers from the Zeno effect,
exactly like the analogous expression Eq.(\ref{2.9}) (in the limit $\epsilon \rightarrow 0 $)
in the non-relativistic arrival time problem. But the key idea here is that we may also get around the problem
in the same way, using a complex potential.
That is, we ``soften'' the restricted propagator and make the replacement
\beq
g_r (t_2, t_1)  \rightarrow \exp \left( - i H (t_2 - t_1) - V ( t_2 - t_1) \right)
\eeq
where $V (q ) = V_0 f_{\Delta} ( q ) $. Here, $V_0 > 0 $ is a constant and $f_{\Delta} (q )$
is the characteristic function of $\Delta$, so is $ 1$
in $\Delta $ and $ 0 $ outside it. Or we may equivalently write $ V = V_0 P $, where recall, $P$
is the projector onto $\Delta$.
This means that our new definition for the class operator is
\beq
\bar C_{ \Delta} = \lim_{t_2 \rightarrow \infty, t_1 \rightarrow -\infty}
\  e^{ i H t_2} \ \exp \left( - i (H  - i V )(t_2 - t_1)  \right) \ e^{ - i H t_1}
\label{C10}
\eeq
The class operator thus defined has an appealing form: it is the S-matrix for system
with Hamiltonian Eq.(\ref{1.3}) scattering off the complex
potential $ V $. As required, it commutes with $H$. This is the most important definition of the paper.

Now an important observation. The class operator derived above has been defined as
a time-ordered product of operators in which the direction of parameter time
increases from right to left. However, since parameter time is unphysical, there
is absolutely no reason why the parametrization should not run in the opposite
direction. This produces an operator which is the hermitian conjugate of Eq.(\ref{C10}).
As one can see from Eq.(\ref{1.20}), this makes no difference in the final expressions
for probabilities. In fact, it is most natural to define the class operator in such
a way that it is invariant under reversing the direction of parametrization. We
thus define a modified class operator which is hermitian:
\beq
\bar C'_{\Delta} = \half \left( \bar C_{\Delta} + \bar C_{\Delta}^{\dag} \right)
\label{Cmod}
\eeq
In what follows, for simplicity, we will primarily work with the non-hermitian
class operator Eq.(\ref{C10}) and revert to the hermitian one, Eq.(\ref{Cmod})
where appropriate. The difference between them will turn out to be significant only
for the class operator for two or more regions.

We now cast the above class operator in a more useable form. The following identities are readily
derived:
\bea
e^{ - i (H - i V) (t_2 - t_1) } &=& e^{ - i H  (t_2 - t_1) } -
\int_{t_1}^{t_2} dt \ e^{ - i H (t_2 - t) } V e^{ - i ( H - i V ) (t- t_1 ) }
\\
&=& e^{ - i H  (t_2 - t_1) } -
\int_{t_1}^{t_2} dt \ e^{ - i (H - i V) (t_2 - t) } V e^{ - i H  (t- t_1 ) }
\eea
Inserting the second expression in the first, we obtain
\bea
e^{ - i (H - i V) (t_2 - t_1) } &=& e^{ - i H  (t_2 - t_1) } -
\int_{t_1}^{t_2} dt \ e^{ - i H (t_2 - t) } V e^{ - i H  (t- t_1 ) }
\nonumber \\
&+& \int_{t_1}^{t_2} dt \int_{t_1}^t ds \ e^{ - i H (t_2 - t) } V e^{ - i ( H - i V) (t-s) }
V e^{ - i H (s - t_1) }
\label{C13}
\eea
Inserting in the expression for the class operator (\ref{C10}) and taking the limit, we obtain
\bea
\bar C_{\Delta} &=& 1 - \int_{-\infty}^{\infty} dt \ V(t)
\nonumber \\
&+& \int_{-\infty}^{\infty} dt \int_{-\infty}^t ds \ e^{  i H t } V e^{ - i ( H - i V) (t-s) }
V e^{ - i H s  }
\label{C14}
\eea
The class operator for entering the region is therefore given by
\beq
C_{\Delta} = \int_{-\infty}^{\infty} dt \ V(t)
- \int_{-\infty}^{\infty} dt \int_{-\infty}^t ds \ e^{  i H t } V e^{ - i ( H - i V) (t-s) }
V e^{ - i H s  }
\label{C15}
\eeq
This is an exact and useful form for the class operator for entering $\Delta$, and it is easily
confirmed that it is of the form Eq.(\ref{1.11}) so commutes with $H$.

Now we use a simple but useful semiclassical approximation. Noting that $ V = V_0 P $, where
$P$ recall projects into $\Delta$, note that the expression
\beq
V e^{ - i ( H - i V) (t-s) } V
\label{semi1}
\eeq
describes propagation with the complex Hamiltonian $ H - i V $ between two points that lie
inside $\Delta$. This can easily be represented by a path integral in which it
seems plausible that the dominant paths between this two end-points will lie entirely
inside $\Delta$, as long
as the boundary is reasonable smooth and the semiclassical paths are not too irregular. If this is true,
we may replace $ V = V_0 P $ by the constant complex potential $V = V_0$,
that is,
\beq
V e^{ - i ( H - i V) (t-s) } V \approx V e^{ - i ( H - i V_0) (t-s) } V
\label{semi}
\eeq
Propagation with a complex potential in Eq.(\ref{semi1}) will also
involve reflection off the boundary of the region, in which case
the semiclassical approximation Eq.(\ref{semi}) may fail.
However, reflection is small for sufficiently small $V_0$ \cite{HaYe1,complex},
and we will see this in more detail in Section 5. Hence we expect te semiclassical
approximation Eq.(\ref{semi}) to hold for small $V_0$.

With this useful approximation (which does not affect the fact that the class operator
commutes with $H$), we have
\beq
C_{\Delta} = \int_{-\infty}^{\infty} dt \ V(t)
- \int_{-\infty}^{\infty} dt \int_{-\infty}^t ds \ V(t) V(s) e^{ - V_0 (t-s) }
\eeq
Again using $ V = V_0 P$, this is easily rewritten
\beq
C_{\Delta} = \int_{-\infty}^{\infty} dt \ P(t) \int_{-\infty}^t ds  \ V_0 e^{ - V_0 (t-s)}
\dot P(s)
\label{C19}
\eeq
This is the main result of the paper: a class operator commuting with $H$, describing
histories which enter the region $\Delta$. It is valid in the approximation Eq.(\ref{semi}), which is sufficient
to cover the key case of histories which, classically, intersect the boundary of $\Delta $ twice.

For systems whose classical paths intersect the surfaces of interest more than two times, the semiclassical approximation Eq.(\ref{semi}) will not be valid, but
the exact result Eq.(\ref{C15}) may still be used. It may also be of interest to explore
higher order semiclassical approximations obtained by iterations of the basic result Eq.(\ref{C13}).
This will not be explored here.

\subsection{Class Operators for Two Regions}

We now extend the above results to the case of histories which enter two regions,
$\Delta_1, \Delta_2$ where we define $P_1$ and $P_2$ to be the projection operators
onto these regions. We proceed as follows.
The class operator Eq.(\ref{C14})
is defined for any complex potential $V$ so we may use a potential
\bea
V &=& V_1 + V_2
\nonumber \\
&=& V_{10} P_1 + V_{20} P_2
\label{pot}
\eea
which has support only in the regions $\Delta_1, \Delta_2$. Here $V_{10}$ and $V_{20}$
are positive constants.
This potential inserted in Eq.(\ref{C14})
defines the class operator $ \bar C_{\Delta_1 \Delta_2} $
for histories which remain outside of both $\Delta_1 $ and $\Delta_2$.
The negation of this class operator therefore describes histories which enter $\Delta_1$ or $\Delta_2$,
or both, and we may write
\beq
\bar C_{\Delta_1 \Delta_2} = 1 - C_{\Delta_1} - C_{\Delta_2} - C_{\Delta_1 \Delta_2}
\eeq
Here, $C_{\Delta_1}$ and $C_{\Delta_2}$ are the class operators for entering the regions
$\Delta_1$, $\Delta_2$ and are given by expressions of the form Eq.(\ref{C15}). We may thus deduce
the form of the class operator $C_{\Delta_1 \Delta_2}$ for entering both regions.

The class operator
Eq.(\ref{C14}) with the potential Eq.(\ref{pot}) is
\bea
\bar C_{\Delta_1 \Delta_2}
&=& 1 - \int_{-\infty}^{\infty} dt \ (V_1(t) + V_2 (t) )
\nonumber \\
&+& \int_{-\infty}^{\infty} dt \int_{-\infty}^t ds \ e^{  i H t } (V_1 + V_2) e^{ - i ( H - i (V_1 +V_2)) (t-s) }
(V_1  + V_2) e^{ - i H s  }
\eea
It seems reasonable to make the semiclassical approximations
\bea
V_1  e^{ - i ( H - i (V_1 +V_2)) (t-s) } V_1 & \approx & V_1  e^{ - i ( H - i V_1 ) (t-s) } V_1
\label{3.23} \\
V_2  e^{ - i ( H - i (V_1 +V_2)) (t-s) } V_2 & \approx & V_2  e^{ - i ( H - i V_2 ) (t-s) } V_2
\label{3.24}
\eea
These approximations are very similar to Eq.(\ref{semi}). Eq.(\ref{3.23})
is essentially the notion that the propagator between two points in $\Delta_1$ will involve negligible
contribution from paths that
enter $\Delta_2$. Similarly for Eq.(\ref{3.24}).
With these approximations we can easily identify the terms $C_{\Delta_1}$ and $C_{\Delta_2}$
and we deduce
\bea
C_{\Delta_1 \Delta_2} = &-& \int_{-\infty}^{\infty} dt \int_{-\infty}^t ds \ e^{  i H t } V_2 e^{ - i ( H - i (V_1 +V_2)) (t-s) }
V_1  e^{ - i H s  }
\nonumber \\
&-& \int_{-\infty}^{\infty} dt \int_{-\infty}^t ds \ e^{  i H t } V_1 e^{ - i ( H - i (V_1 + V_2)) (t-s) }
V_2  e^{ - i H s  }
\label{3.25}
\eea

To estimate the propagator
\beq
V_2 e^{ - i ( H - i (V_1 +V_2)) (t-s) } V_1
\eeq
we need a semiclassical approximation more elaborate than Eq.(\ref{semi}), since it involves
propagation from  an initial point in $\Delta_1$ with potential $V_1$, to  a final point in $\Delta_2$ with
potential $V_2$ but with zero complex potential outside these regions. The path decomposition
expansion \cite{PDX,HaOr,Hal3} is naturally adapted to this problem, since it is a tool for breaking up
a propagator into its properties in different spatial regions. We will simply write down
the result of using this expansion, with some brief justification.
\bea
P_2 e^{ - i ( H - i (V_1 +V_2)) (t-s) } P_1 \approx
\int_{s}^t ds' \int_{s'}^t & dt' & \ P_2 e^{ - i H (t-t') } \dot P_2 e^{ - i H (t'-s') } \dot P_1
e^{ - i H (s'-s) } P_1 \
\nonumber \\
& \times & \exp \left( - V_{20} (t-t') - V_{10} (s'-s) \right)
\eea
This expression is actually exact when the complex potential is absent, as is easily verified.
Since $ \dot P_1 $ and $\dot P_2 $ represent the fluxes at the boundary of $ \Delta_1 $ and
$\Delta_2$, one can easily identify the sections of the propagation which, semiclassically,
are in $\Delta_1$, outside $\Delta_1 $ and $\Delta_2$, and  in $\Delta_2$. The exponential
suppression factors involving the complex potential for propagation inside $\Delta_1 $ and $\Delta_2$
are inserted using a semiclassical approximation
along the lines of Eq.(\ref{semi}). Inserting in Eq.(\ref{3.25}), we thus obtain
\bea
C_{\Delta_1 \Delta_2} = - \int_{-\infty}^{\infty}   dt  \int_{-\infty}^t & ds & \int_{s}^t ds' \int_{s'}^t dt'
\ V_2 (t) \dot P_2 (t') \dot P_1 (s') V_1 (s)
\nonumber \\
& \times & \exp \left( - V_{20} (t-t') - V_{10} (s'-s) \right)
\nonumber \\
 - \int_{-\infty}^{\infty}  dt  \int_{-\infty}^t & ds & \int_{s}^t ds' \int_{s'}^t dt'
\ V_1 (t) \dot P_1 (t') \dot P_2 (s') V_2 (s)
\nonumber \\
& \times & \exp \left( - V_{10} (t-t') - V_{20} (s'-s) \right)
\label{C30}
\eea

\section{Properties of the Class Operators}

We now examine the properties of the class operators Eq.(\ref{C19}), (\ref{C30}) and confirm that they give the desired
results.

\subsection{Single Region}

It is easy to show that the class operator for a single region Eq.(\ref{C19}) has a sensible classical limit.
Classically, $P(t)$ is
a function on classical trajectories with $P(t) = 1 $ when the classical trajectory
is in $\Delta$ and is zero otherwise. Suppose a given trajectory $q^a (t)$ enters $\Delta$
at some stage in its history so intersects the boundary twice (recalling we have essentially assumed
no more than two intersections in the semiclassical approximation Eq.(\ref{semi})).
For a given choice of parametrization of the trajectory, it
enters $ \Delta $ at parameter
time $t_a$ and
leaves at time $t_b > t_a$ (see Figure 4). (We may parameterize it in
the opposite direction, with the same ultimate result).
Then the derivative of $P(s)$ is
\beq
\dot P(s) = \delta ( s - t_a) - \delta ( s - t_b)
\eeq
and we have
\beq
C_{\Delta} = \int_{t_a}^{t_b} dt \ \int_{t_a}^t ds  \ V_0 e^{ - V_0 (t-s)}
\left[ \delta ( s - t_a) - \delta ( s - t_b) \right]
\label{P2}
\eeq
Since $ s \le t \le t_b $, the second $\delta$-function in Eq.(\ref{P2}) makes no contribution
to the integral. This is exactly the desired property -- the expression for $C_{\Delta}$ registers
only the first intersection of the trajectory with the boundary, but not the second intersection.
The integral is easily evaluated with the result,
\beq
C_{\Delta} = 1 - e^{ - V_0 (t_b - t_a) }
\label{P3}
\eeq
This is approximately $1$, as required, as long as
\beq
V_0 ( t_b - t_a ) \gg 1
\eeq

We now give the broad picture in the quantum case and confirm some of the details
in some simple models in the next section.
Suppose we operate with $C_{\Delta}$ on an eigenstate $ |\Psi_\l \rangle$
of $H$. The time integrals in Eq.(\ref{C19}) are easily carried out and the result is
\beq
C_{\Delta} | \Psi_\l \rangle = 2 \pi V_0 \delta (H - \lambda ) \ P G_V \dot P | \Psi_\l \rangle
\label{P4}
\eeq
where
\bea
G_V &=& \int_0^{\infty} dt \ e^{  - i (H - \l)t - V_0 t}
\label{P6a}\\
&=&\frac { - i } { H - \l - i V_0 }
\label{P6}
\eea
and we have used
\beq
\delta ( H - \l) = \frac {1} { 2 \pi} \int_{-\infty}^{\infty} dt \ e^{  - i (H - \l)t }
\label{P8}
\eeq
The $\dot P$ term is a current operator on the boundary $\Sigma$ of $\Delta$ and may be written
\bea
\dot P &=& i [ H, P ]
\nonumber \\
&=& - \hat p_n \delta_{\Sigma} (\hat q) - \delta_{\Sigma} (\hat q) \hat p_n
\eea
where
\beq
\delta_{\Sigma} (\hat q) =  \int_{\Sigma} d^{n-1} q \ | q  \rangle \langle q |
\eeq
is a $\delta$-function operator in the surface $\Sigma$ and
$\hat p_n = n^a \hat p_a $ is the
component of the momentum operator normal to $\Sigma $
with $n^a$ the outward pointing normal. The normal $n^a$ will depend on $q$ in general
so there may be an operator ordering issue in making $\hat p_n$ hermitian. The difference between different ordering will involve a term of the form $\nabla_a n^a $, essentially
the extrinsic curvature of $\Sigma$, and this will be small if $\Delta$ is sufficiently
large and its boundary reasonably smooth.
Note that it is sometimes also convenient to write $\dot P$ as
\beq
\dot P
=  i \int_{\Sigma} d^{n-1} q \ | q \rangle \overleftrightarrow {\partial_n} \langle
q |
\label{4.11a}
\eeq

It is very useful to separate the current operator
at the boundary into ingoing and outgoing parts according to the sign of $\hat p_n$
at the boundary
\beq
\dot P = ( \dot P)_{in} - ( \dot P)_{out}
\label{pdot}
\eeq
where $ (\dot P)_{in}$ consists of the components of $ \dot P$ with incoming momentum
\beq
(\dot P)_{in} = - \hat p_n \theta ( - \hat p_n) \delta_{\Sigma} (\hat q) - \delta_{\Sigma} (\hat q) \hat p_n \theta ( - \hat p_n)
\eeq
and similarly for $ (\dot P)_{out}$.
Examples of these expressions in particular models will be given in the following sections. The restriction to positive or negative $p_n$ means that it is generally
difficult to express $(\dot P)_{in}$ and $(\dot P)_{out}$ in the form Eq.(\ref{4.11a}),
involving the derivative $ \hat p_n = - i \partial_n$, unless operating on a state
(such as a WKB state) with simplifying properties.
Note that these definitions require only that the local flux operator on a {\it given} surface
can be split into ingoing and outgoing parts. To do so globally on a family of surfaces
is generally impossible, essentially due to the problem of time, but fortunately
this is not required here.

The quantity
$G_V$ has the form of a Feynman propagator, Eq.(\ref{GF}),
with $V_0$ playing the role of the ``$ i \epsilon$
prescription", as long as $V_0$ is sufficiently small (in comparison to an appropriate energy scale
contained in $H$, such as $p_0$ in the case of the Klein-Gordon equation). We therefore
expect it to have properties similar to $G_F$, as described in Section 1D.

However, $V_0 $ is not set to zero exactly and this in fact
means that $G_V$ has two properties not possessed by $G_F$.
First, the non-zero $V_0$ produces a suppression for widely
separated initial and final points.
In a path integral representation of $ G_V (q,q')$, the sum will be dominated by
classical paths from $q$ to $q'$, to which one may associate the total parameter
time $\tau $ of the path. From the integral representation Eq.(\ref{P6a})
it can be seen that $G_V$ will have an overall exponential suppression factor
of the form $\exp ( - V_0 \tau )$. Thus, propagation with $G_V$ will suppress
configurations $q$ and $q'$ connected by classical trajectories of parameter time
duration of greater than $1/ V_0$. Second, recalling that the class operator
is closely related to the $S$-matrix for scattering off a complex potential
there will be some reflection involved and this appears in properites of $G_V$ -- it will not exactly possess the Feynman properties of propagating positive
frequencies to the future negative frequencies to the past, but may,
for example, propagate some positive frequencies to the past. However,
this ``non-Feynman'' propagation will be small for sufficiently small
$V_0$, as we will see in the next section, so this possibility will be
ignored.

These properties of $G_V$ are crucial to understanding the properties of $C_{\Delta}$.
First note that
\beq
\langle q | P G_V \dot P | \Psi_\l \rangle
= - i \int_{\Sigma} d^{n-1} q' \ f_{\Delta} (q) G_V (q,q') \overleftrightarrow {\partial_n}  \Psi_{\l} (q')
\eeq
In this expression the propagator $G_V (q,q') $ propagates from the
boundary $\Sigma$ to a point $q$ on the interior.
The fact that $G_V$ is, approximately, a propagator of
the Feynman type means two things. First, there will be initial and final points $q$, $q'$
for which the propagator is very small (due to ``propagation outside the lightcone'',
discussed in Section 1D).
Secondly, when it is not small, one would expect it to involve only
{\it ingoing} modes at the boundary (to the extent that reflection is ignored).
Hence the $P G_V$ terms effectively restrict the
current operator $\dot P$ at the boundary to ingoing modes only -- exactly the result we are
looking for. We therefore replace $ \dot P$, with the current operator $ ( \dot P )_{in} $ involving
ingoing modes only
\beq
C_{\Delta} | \Psi_\l \rangle = 2 \pi V_0 \delta (H - \lambda ) \ P G_V ( \dot P )_{in} | \Psi_\l \rangle
\eeq

Now note that since the $P G_V$ terms have done their job of selecting the ingoing modes,
they may be eliminated. More precisely, we write
$ P = 1 - \bar P$, where $\bar P$ is the projector onto the region outside $\Delta$, and noting that
that
\beq
V_0 \delta ( H - \l ) G_V = \delta ( H - \l)
\label{P13}
\eeq
we find
\bea
C_{\Delta} | \Psi_\l \rangle &=& 2 \pi \delta ( H - \l) ( \dot P )_{in} | \Psi_\l \rangle
\nonumber \\
&-& 2 \pi \delta ( H - \l ) \bar P G_V ( \dot P )_{in} | \Psi_\l \rangle
\label{P14}
\eea
Consider the second term on the right-hand side. It consists of ingoing modes at the boundary
of $\Delta$ propagated with $G_V$ to a final point which is {\it outside} $\Delta$. Semiclassically,
this corresponds to paths which have to traverse the entire width of $\Delta$, so if $\Delta$
is sufficiently large, this will take a long parameter time $\tau$ and, as discussed above,
$G_V$ will contain a suppression
factor of $\exp ( - V_0 \tau) $ in comparison to the first term in Eq.(\ref{P14}). This is exactly
analogous to the suppression of the second term in the classical case, Eq.(\ref{P3}).
We thus deduce that
\beq
C_{\Delta} | \Psi_\l \rangle \approx 2 \pi \delta ( H - \l) ( \dot P )_{in} | \Psi_\l \rangle
\label{P15}
\eeq
Note that the approximation leading to dropping the second term in Eq.(\ref{P14})
also ensures that the result is independent of $V_0$, like the classical case in the appropriate
limit. Eq.(\ref{P15}) is the main result of this section.

Eq.(\ref{P15}) may also be expressed in terms of $(\dot P)_{out}$ defined in Eq.(\ref{pdot}).
To see this, note that, since $ \dot P = i [H,P]$, we have
\beq
\delta ( H - \l') \dot P | \Psi_\l \rangle = i ( \l' - \l ) \delta ( H - \l')  P | \Psi_\l \rangle
\eeq
This is zero for $ \l = \l' $, as long as $ \langle \l | P | \l \rangle $ is well-defined,
where $ |\l  \rangle $ are eigenstates of $H$. (It may not be well-defined
if $P$ projects onto an infinite region -- see Appendix A). Eq.(\ref{pdot}) then implies
\beq
\delta ( H - \l) ( \dot P )_{in} | \Psi_\l \rangle
= \delta ( H - \l) ( \dot P )_{out} | \Psi_\l \rangle
\label{P17a}
\eeq
so Eq.(\ref{P15}) may be expressed in terms of either the ingoing or outgoing flux
or both.

It is also useful to note that the right-hand side of Eq.(\ref{P15}) may be written
\beq
2 \pi \delta (H - \l ) (\dot P)_{in} | \Psi_\l \rangle =  I_{\Sigma} | \Psi_\l \rangle
\eeq
where
\beq
I_\Sigma  = \int_{-\infty}^{\infty} dt \ e^{ i H t } (\dot P )_{in} e^{ - i H t }
\label{int}
\eeq
This is an intersection number operator for ingoing flux at the boundary $\Sigma$ of $\Delta$.
Hence the class operator is essentially the intersection number $I_{\Sigma}$,
and clearly commutes with $H$.
It is classically equal to $1$ for trajectories with ingoing flux at $\Sigma$
(with no more than two intersections of the boundary, in the approximation
we are using),
and zero for trajectories not intersecting $\Sigma$. Of course, one may have
guessed this approximate formula for the class operator, but a class operator is fundamentally
defined as a product of projectors (or quasi-projectors) and the derivation given here makes
it clear how this obvious guess arises from the fundamental definition.
Note also the the class operator is hermitian in this case, so there is no need to consider
the modified propagator Eq.(\ref{Cmod}).

The form Eq.(\ref{P15}) may be used to check for decoherence of histories in specific models
and we will see this later.
When there is decoherence of histories, the probabilities are given by the average of
a single class operator, Eq.(\ref{1.20}), which in this case reads
\beq
\langle \Psi_{\l'} | C_{\Delta} | \Psi_\l \rangle
= 2 \pi \langle \Psi_{\l} |( \dot P )_{in} | \Psi_\l \rangle \ \delta (\l - \l')
\eeq
Following the induced inner product prescription, we drop the $\delta$-function
on the right and then set $\l=\l' = 0 $, so the probability in terms of
a solution $| \Psi \rangle $ of the Wheeler-DeWitt equation is
\beq
\langle \Psi | C_{\Delta} | \Psi \rangle_{phys}
= 2 \pi \langle \Psi |( \dot P )_{in} | \Psi \rangle
\label{P17}
\eeq
This is essentially the ingoing Klein-Gordon flux on the boundary of $\Delta$,
as expected. (The factor of $ 2 \pi$ relates
to the induced inner product as described in Appendix A).

A more detailed calculation in a specific model is required to see that the above argument
works. In particular,
it is necessary to show that the various requirements on $V_0$ can be simultaneously met --
it has to be small enough for $G_V$ to function as a Feynman propagator and for
reflection to be neglected,
but large enough to ensure independence of $V_0$ and the dropping of the second term in Eq.(\ref{P14}).
We will see in the models below that this is indeed possible.

\subsection{Two Regions}

We first consider the class operator Eq.(\ref{C30}) for two regions in the classical case.
Suppose a classical trajectory enters $\Delta_1$ at $t_a$, leaves at $t_b$, enters
$\Delta_2$ at $t_c$ and leaves at $t_d $ (with no more than two crossings of the boundaries
of each region). See Figure 5. Then we have
\beq
\dot P_2 (t') \dot P_1 (s') = \left[ \delta (t'-t_c) - \delta ( t'- t_d ) \right]
 \left[ \delta (s'-t_a) - \delta ( s'- t_b ) \right]
\eeq
It is easy to see that only one of the two terms in Eq.(\ref{C30})
contributes -- the other term corresponds to the time-reversed trajectory.
One can also see that the $\delta$-functions at
$t'=t_d$ and $s'=t_a$ make no contribution to the integral.
Hence the only contribution
comes from the time of departure $t_b$ from $\Delta_1$ and the time $t_c$ of
entering $\Delta_2$, clearly sufficient to characterize histories which spend time in both regions without overcounting.
Evaluating the integral we find
\beq
C_{\Delta_1 \Delta_2} = \left( 1 - e^{- V_{20} (t_d - t_c) } \right)
\ \left( 1 - e^{- V_{10} (t_b - t_a) } \right)
\eeq
which is approximately $1$, as required, as long as the trajectory spends sufficiently
long in each region. For trajectories entering only one region, the $\dot P$ term
associated with the other region is zero, so the class operator is zero.
Classically, the class operator therefore has all the desired properties.

In the quantum case, we follow steps similar to the case for one region considered
above. We consider the action of the class operator $C_{\Delta_1 \Delta_2 }$, defined
in Eq.(\ref{C30}) on an eigenstate $ | \Psi_\l \rangle $
of $H$. Changing variables from $s, s',t'$ to
\bea
\bar s &=& t - s
\nonumber \\
\bar s' &=& s' -s
\nonumber \\
\bar t' &=& t - t'
\eea
the $t$ integral may be done with the result
\bea
C_{\Delta_1 \Delta_2 } | \Psi_\l \rangle = &-& 2 \pi  \int_0^\infty d \bar s
\int_0^{\bar s}  d \bar s'   \int_0^{\bar s - \bar s'} d \bar t' \ \exp \left(  -V_{20} \bar t' - V_{10} \bar s' \right)
\nonumber \\
& \times & \ \delta (H - \l ) V_2 e^{ - i H \bar t'} \dot P_2 e^{ - i H ( \bar s
- \bar t' - \bar s') } \dot P_1 e^{ - i H \bar s'} V_1 | \Psi_\l
\rangle e^{ i \l \bar s }
\nonumber \\
&+& (1 \ \leftrightarrow \ 2)
\eea
Now note that
\bea
\dot P_1 e^{-i H \bar s'}  V_1 | \Psi_\l \rangle
 & \approx & - V_{10}
(\dot P_1)_{out} e^{-i H \bar s'} P_1 | \Psi_\l \rangle
\nonumber \\
&  \approx  & - V_{10} e^{ - i \l \bar s'}
(\dot P_1)_{out}  | \Psi_\l \rangle
\eea
The first approximation arises because the combination $ \dot P_1 \exp ( - i H \bar
s') P_1 $ is propagation from inside $\Delta_1$ to the boundary with only positive
$\bar s'$, which means that there are only outgoing modes at the boundary. The second
approximation arises because, once the restriction is made to outgoing modes,
the projection $P_1$ may be dropped (similar to dropping the second term in Eq.(\ref{P14})).
Similar approximations may be made with the terms $ \delta (H - \l ) V_2 e^{ - i
H \bar t'} \dot P_2
$. We thus obtain
\bea
C_{\Delta_1 \Delta_2 } | \Psi_\l \rangle = && 2 \pi  V_{10} V_{20}\int_0^\infty d \bar s
\int_0^{\bar s}  d \bar s'   \int_0^{\bar s - \bar s'} d \bar t' \ \exp \left(  -V_{20} \bar t' - V_{10} \bar s' \right)
\nonumber \\
& \times & \ \delta (H - \l ) (\dot P_2)_{in} e^{ - i (H-\l) ( \bar s
- \bar t' - \bar s') } (\dot P_1 )_{out}| \Psi_\l \rangle
\nonumber \\
&+& (1 \ \leftrightarrow \ 2)
\eea
The integrals may be carried out with the result,
\beq
C_{\Delta_1 \Delta_2 } | \Psi_\l \rangle  =  2 \pi \delta ( H - \l)
\left[  (\dot P_2)_{in} G_F (\l) ( \dot P_1)_{out} + (1 \ \leftrightarrow \ 2) \right] | \Psi_\l \rangle
\label{P25}
\eeq
where $G_F (\l)$ is the Feynman propagator at eigenvalue $\l$,
\bea
G_F (\l ) &=& \int_0^\infty dt \ e^{ - i ( H - \l - i \epsilon ) }
\nonumber \\
&=& \frac {-i} { H - \l - i \epsilon}
\eea
We may again attempt to interchange ingoing and outgoing flux operators, along
the lines of Eq.(\ref{P17a}). However, since
\beq
(H - \l) G_F (\l) = - i
\eeq
there are extra terms of the form $ (\dot P_2)_{in} P_1 $
and $( \dot P_1)_{in} P_2 $, but these are zero since they are products
of operators localized about spatially distinct regions. We thus find
\beq
C_{\Delta_1 \Delta_2 } | \Psi_\l \rangle  =  2 \pi \delta ( H - \l)
\left[  (\dot P_2)_{in} G_F (\l) ( \dot P_1)_{in} + (1 \ \leftrightarrow \ 2) \right] | \Psi_\l \rangle
\label{P34}
\eeq
(or equivalently in terms of outgoing flux).

At this stage it is of interest to examine the difference between this propagator
and the modified version, Eq.(\ref{Cmod}), invariant under reversals of parameter
time, since the presence of $G_F (\l)$ in Eq.(\ref{P34}) rather than $ \delta (H - \l )$ means that it is not invariant. It is easy to see that the modification
Eq.(\ref{Cmod}) effectively produces the replacement
\beq
G_F (\l) \rightarrow \half \left( G_F (\l ) + G_F^{\dag} (\l) \right)
= \half (2 \pi) \delta ( H - \l)
\label{P35}
\eeq
and we obtain for the modified propagator
\bea
C'_{\Delta_1 \Delta_2 } | \Psi_\l \rangle  & =  & \half (2 \pi)^2 \delta ( H - \l) (\dot P_2)_{in} \delta (H - \l) ( \dot P_1)_{in} | \Psi_\l \rangle
\nonumber \\
& + & \half (2 \pi)^2 \delta ( H - \l) (\dot P_1)_{in} \delta (H - \l) ( \dot P_2)_{in} | \Psi_\l \rangle
\label{P36}
\eea
This result may also be expressed in terms of intersection number
operators,
\beq
C'_{\Delta_1 \Delta_2} =
\half \left( I_{\Sigma_2}  I_{\Sigma_1}  + I_{\Sigma_1} I_{\Sigma_2} \right)
\eeq
It is now easy to guess the form of the class operator for $n$ regions. It is,
\beq
C'_n = \frac {1} {n!} \left( I_{\Sigma_1} I_{\Sigma_2} \cdots I_{\Sigma_n} + \ permutations
\right)
\eeq
where ``permutations'' means add all possible permutations of the $n$ regions, to
give a total of $n!$ terms. This final result has a particularly natural form
which also suggests that it may have simple path integral representations.
This will be explored elsewhere.
(See also Ref.\cite{Haltraj} for similar expressions, derived using a detector model
for quantum cosmology).

\section{Some Simple Examples}

To verify the properties of the class operator $C_{\Delta}$ outlined above, we compute
it explicitly in some simple examples. We first consider a one-dimensional example.

\subsection{A One-Dimensional Example}

We take the Hamiltonian
to be the free particle Hamiltonian $H = p^2 / 2m$ and the region $\Delta$ to be
the interval $ [ -L, L] $. We take the initial state to be an energy eigenstate
\beq
\langle x | p \rangle = \frac {1} { ( 2 \pi)^{1/2} } e^{ i p x}
\eeq
with $ p > 0 $ and energy $ E = p^2 / 2m $. (In this case, unlike the Wheeler-DeWitt
and Klein-Gordon equations, the eigenvalue of $H$ is the physical energy, so we do
not take $E = 0$ at the end of the calculation. However, the inner products
are still regularized by taking different values of $E$ on either side of any inner
product.)

Our aim is to confirm in this example the properties of the class operator Eq.(\ref{P4}),
which in this case we write as,
\beq
C_{\Delta} | p \rangle = 2 \pi V_0 \delta ( H - E) P G_V \dot P | p \rangle
\eeq
$P$ is the projection operator onto the region $[-L,L]$ and may also be
represented in terms of the region's characteristic function $f_L (x)$,
so $ P = f_L ( \hat x ) $. It follows that
\beq
\dot P = \frac { 1} {2m} \left[
\hat p \delta ( \hat x + L ) + \delta ( \hat x + L ) \hat p
-\hat p \delta ( \hat x - L ) - \delta ( \hat x - L ) \hat p \right]
\eeq
For the $p>0$ initial state considered here, we identify the first two terms,
at $x=-L$,
as $(\dot P)_{in}$ and the last two terms, at $x=L$, as $ -(\dot P)_{out}$,
\beq
\dot P = ( \dot P)_{in} - ( \dot P)_{out}
\eeq
We may now write
\bea
\langle  x | P G_V \dot P | p \rangle &=&
\int dy \ G_V (x,y) \langle y | \dot P | p \rangle
\nonumber \\
&=& \frac { -i} {2m  \sqrt{2 \pi} }  f_L (x) \left(   \left[
G_V (x,y) \overleftrightarrow {\partial_y} e^{ i p y} \right]_{y=-L}
- \left[ G_V (x,y) \overleftrightarrow {\partial_y} e^{ i p y}\right]_{y=L} \right)
\label{5.5}
\eea
The Green function may be calculated explicitly using the form Eq.(\ref{P6a}), which
may be written
\beq
G_V(x,y) = \int_0^{\infty} dt \ \left( \frac {m} {2 \pi i t} \right)^{1/2}
\ \exp \left(  i m \frac{ (x-y)^2 } {2 t} + i ( E + i V_0 ) t \right)
\eeq
The integral may be carried out explicitly \cite{Sch}, with the result
\beq
G_V(x,y) = \frac { m } {Q} \left[ \theta (x-y) e^{ i (x-y) Q}
+ \theta (y-x) e^{ - i (x-y) Q } \right]
\eeq
where we have introduced the complex momentum
\beq
Q = \sqrt{ 2m (E + i V_0)}
\eeq
The interesting case is that in which $V_0 \ll E $ (which is the condition
for negligible reflection in the scattering off the complex potential) and we can then
expand
\bea
Q &=& \sqrt{ 2m E} \left( 1 + i \frac {V_0} {E} \right)^{1/2}
\nonumber \\
&=& p + i \frac {V_0 m } {p} + \cdots
\eea
where $ p = \sqrt{ 2 m E } $.
This means that the Green function has the form
\beq
G_V(x,y) \approx \frac { m } {p} \left[ \theta (x-y) e^{ i (x-y) p}
+ \theta (y-x) e^{ - i (x-y) p } \right] \ \exp \left( - \frac {V_0 m } {p} | x - y | \right)
\eeq
It therefore has the form of Feynman type Green function, but there is also exponential
suppression for large separations of initial and final points, as anticipated.

To evaluate Eq.(\ref{5.5}), we consider the action of the Green function $G_V$ on
a plane wave state $ | p \rangle $ with $p>0$. We have
\bea
\frac {-i} {2m} G_V (x,y) \overleftrightarrow {\partial_y} e^{ i p y} &=&
\theta (x-y) \frac { (Q+p) } { 2 Q} \exp \left( i p x + i (x-y) (Q-p) \right)
\nonumber \\
&-& \theta ( y - x) \frac { (Q-p)} { 2 Q} \exp \left( - i p x + 2 i p y + i (y-x) (Q - p) \right)
\eea
For $ E \gg V_0$, this has the approximate form,
\bea
\frac {-i} {2m} G_V (x,y) \overleftrightarrow {\partial_y} e^{ i p y} &=&
\theta (x-y) \ e^{ i p x} \ \exp \left( - \frac {V_0 m } {p} | x - y | \right)
\nonumber \\
&-& 4 i  \frac { V_0} {E} \theta ( y - x) \ e^{ - i p x + 2 i p y}
\ \exp \left( - \frac {V_0 m } {p} | x - y | \right)
\eea
The first term corresponds to the familiar property of consisting of positive momentum
for $x>y$ (with the exponential suppression factor mentioned above). The second term
is not a familiar property of the Feynman Green function since it corresponds
to negative momentum for $x<y$ (which we do not expect since the incoming state
consists only of positive momentum). This actually corresponds to reflection, and
this term will be small for $V_0 \ll E $, which we assume.

These results show that the second term in Eq.(\ref{5.5}), the term corresponding to $(\dot P)_{out}$
may be dropped, as expected. We have now confirmed that
\beq
C_{\Delta} | p \rangle \approx 2 \pi V_0 \delta ( H - E) P G_V (\dot P)_{in} | p \rangle
\eeq
Following the general discussion of Section 4, we write $ P = 1 - \bar P $ and obtain
\bea
C_{\Delta} | p \rangle & \approx & 2 \pi \delta ( H - E) (\dot P)_{in} | p \rangle
\nonumber \\
&-& 2 \pi V_0 \delta ( H - E) \bar P G_V (\dot P)_{in} | p \rangle
\eea
We need to show that the second term is small. We have
\beq
\langle x | \bar P G_V (\dot P)_{in} | p \rangle
= \frac { -i} {2m  \sqrt{2 \pi} } \bar f_L (x)  \left[
G_V (x,y) \overleftrightarrow {\partial_y} e^{ i p y} \right]_{y=-L}
\label{5.15}
\eeq
where $ \bar f_L (x) $ is the characteristic function for the region outside
$\Delta $, so $ x \ge L$ and $ x \le - L $. The properties of $G_V$ deduce above
confirm that this expression is small. For $ x \ge L$, there is exponential
suppression of the Green function since $ | x - y | \ge 2 L $. The exponential
suppression is of order $ \exp ( - V_0 \tau ) $, where
\beq
\tau = \frac { 2 m L } { p}
\eeq
which is the time for a classical trajectory to traverse $\Delta$. On the other hand,
values of  $ x \le - L $
can only be reached by reflection, and this is small since $V_0 \ll E $.

We have therefore confirmed the desired result that the class operator has the form
\beq
C_{\Delta} | p \rangle  \approx  2 \pi \delta ( H - E) (\dot P)_{in} | p \rangle
\label{5.17}
\eeq
and this result is valid in the regime
\beq
\frac {1} { \tau} \ll V_0 \ll E
\eeq
(This regime is a commonly encountered one in studies of the arrival time
in non-relativistic quantum mechanics \cite{timeinqm,HaYe1}).
The outer parts of the inequality imply that $ E \tau \gg 1 $, or equivalently,
\beq
p L \gg 1
\eeq
which is easily satisfied for sufficiently large $L$ and $p$.
It is essentially the condition that the size of $\Delta $ is much greater than the
wavelength of the quantum state.

Since
\beq
2 \pi \delta ( H - E) = G_F + G_F^{\dag}
\label{5.19}
\eeq
(where $G_F$ is the Feynman Green function, obtained from $G_V$ by letter $V_0 \rightarrow 0 +$),
we have
\beq
\langle x | \delta ( H - E) | y \rangle = \frac {2 \pi m} {p} \left(
e^{ i p ( x - y )} + e^{ - i p ( x - y )} \right)
\eeq
and it follows that Eq.(\ref{5.17}) may be evaluated with the trivial result
\beq
C_{\Delta} | p \rangle  \approx   | p \rangle
\label{5.20}
\eeq
This is completely expected since every plane wave enters the region $\Delta$.

However, for the class operator operating on the initial state Eq.(\ref{5.17}) to have
a more interesting and non-trivial form, we need to look at a two-dimensional example.
This will be useful for the purposes of examining the decoherence properties of some simple states.

\subsection{A Two-Dimensional Example}

We again take a free particle but in two dimensions, with Hamiltonian
$ H = (p_1^2 + p_2^2)/ 2 m $, and
we take the region $\Delta$ to be a rectangle of sides $2L_1 $ and $2L_2 $ centered at the origin. We use
coordinates $x_1, x_2$ and the characteristic function of $\Delta $ is
\beq
f_{\Delta} ( x_1, x_2 ) = f_{L_1} (x_1) f_{L_2} (x_2)
\eeq
We take the initial state $ | \Psi \rangle$ to be a plane wave in the $1$-direction, $ e^{ i p_1 x_1}$,
with $p_1 > 0 $,
so there is no momentum in the $2$-direction. The analysis of the one-dimensional example
may be used to show, very easily, that the class operator is again of the form Eq.(\ref{5.17}), with
\beq
( \dot P)_{in} = \frac { 1} {2m} \left[
\hat p_1 \delta ( \hat x_1 + L_1 ) + \delta ( \hat x_1 + L_1 ) \hat p_1 \right] \ f_{L_2} ( \hat x_2)
\eeq
Denoting the matrix elements of $ \delta ( H - E)$ by $G (\x| \y)$, we have
\bea
\langle \x | C_{\Delta} | \Psi \rangle &=&
2 \pi \langle x_1, x_2| \delta ( H - E) ( \dot P)_{in} | \Psi \rangle
\nonumber \\&=&
\frac { -i \sqrt {2 \pi} } {2m   } \int_{-L_2}^{L_2} dy_2 \left[
G (x_1, x_2|y_1, y_2) \overleftrightarrow {\partial_{y_1}} e^{ i p_1 y_1} \right]_{y_1=-L_1}
\label{5.24}
\eea
Note that, due to the suppression effect in $G_V$,  the spatial size $2 L_1$ of $\Delta $ in the $x_1$-direction has essentially
dropped out of the expression for the class operator and it is expressed now only
in terms of the incoming flux at $x_1 = - L$.

Eq.(\ref{5.24}) may be approximately calculated, at some length, but the form of the result
for $ C_{\Delta} | \Psi \rangle $ is easy to anticipate on general physical grounds.
We still have the result Eq.(\ref{5.19}), so we first consider the action of $G_F$ attached to
the initial state as in Eq.(\ref{5.24}). The physical situation therefore
consists of an incoming plane wave $\exp ( i p_1 x_1 ) $ from $x_1<0$ encountering a gate
of width $ 2L_2 $ in the $x_2$ direction located at $x_1 = - L$. This situation
is very similar to the Mott analysis of the particle detection in a cloud
chamber \cite{Mott,Haltraj}. The localization by the gate to spatial width $ 2L_2$ produces
a momentum uncertainty in the $2$-direction $ \Delta p_2 $ of order $ 1 / L_2$.
Since the momentum in the $1$-direction is $p_1$, it is easy to see that
this produces a spreading in $x_1>0$ of angular size
of order $ 1/ (p_1 L_2) $. This will be very small for sufficiently large $L_2$.
If we now consider also the action of $G_F^{\dag}$ on the same initial
state, it produces essentially the same effect in $x_1<0$.

The final result is therefore that  $ C_{\Delta} | \Psi \rangle $ consists
of a plane wave $\exp ( i p_1 x_1 ) $ localized on a tube of width $ 2 L_2 $
around the $x_1$-axis, with the tube opening out by a small angle $ 1 / (p_1 L_1)$
as $x_1 \rightarrow \pm \infty $. See Figure 6.
To the extent that the angular spreading can be ignored,
the result has the very approximate form
\beq
\langle x_1, x_2 | C_{\Delta} | \Psi \rangle  \approx  e^{ i p_1 x_1 } f^{\prime}_{L_2} (x_2 )
\label{5.26}
\eeq
where $ f^{\prime}_{L_2} $ is a function localized to a width $L_2$ around the origin
(not necessarily the same as the exact window function used above).

This approximate result may be used to address the decoherence properties of the
initial state $ |\Psi \rangle$ defined above. The class operator for not entering $\Delta$ is clearly
\beq
\langle x_1, x_2 | \bar C_{\Delta} | \Psi \rangle  \approx  e^{ i p_1 x_1 } \bar f^{\prime}_{L_2} (x_2 )
\eeq
where $\bar f^{\prime}_{L_2} = 1 - f^{\prime}_{L_2} $, so $ \bar C_{\Delta} | \Psi \rangle $
is localized outside the tube defined in Eq.(\ref{5.26}). We immediately see that
\beq
\langle \Psi | \bar C_{\Delta} C_{\Delta} | \Psi \rangle \approx 0
\eeq
quite simply because the two class operators are spatially localized about different non-overlapping
regions. There is therefore approximate decoherence of histories in this case.

Another interesting case is to take the same region $ \Delta$ but take a superposition
of initial states
\beq
| \Psi \rangle = \frac {1} {\sqrt{2} } \left( | \Psi_1 \rangle + | \Psi_2 \rangle
\right)
\eeq
where $ | \Psi_1 \rangle $ is a plane wave in the $1$-direction, $\exp (i p_1 x_1 )$, considered
above, and $ | \Psi_2 \rangle $ is a plane wave in the $2$-direction, $\exp ( i p_2 x_2)$. Each
state individually produces decoherence of histories, as shown above, but there will be cross terms
in the decoherence functional,
\beq
\langle \Psi | \bar C_{\Delta} C_{\Delta} | \Psi \rangle \approx
\half \langle \Psi_2 | \bar C_{\Delta} C_{\Delta} | \Psi_1 \rangle
+ \half \langle \Psi_1 | \bar C_{\Delta} C_{\Delta} | \Psi_2 \rangle
\label{5.30}
\eeq
The first term is approximately,
\beq
\langle \Psi_2 | \bar C_{\Delta} C_{\Delta} | \Psi_1 \rangle
\approx \int d x_1 d x_2 \  f^{\prime}_{L_2} (x_2 )  f_{L_1} (x_1 )
\exp \left( i p_1 x_1 + i p_2 x_2 \right)
\eeq
It is easy to see that this expression will be very small as long as
$p_1 L_1 \gg 1 $ and $ p_2 L_2 \gg1 $, due to averaging out of oscillations.
There will therefore be decoherence
of superposition states as long as the region $\Delta$ is sufficiently
large.

The essential reason for decoherence in this case is that the orthogonality
of $ |\Psi_1 \rangle $ and $ | \Psi_2 \rangle $ is little disturbed by the
action of the class operators as long as $\Delta$ is large enough.
(Recall that interference effects in the double slit experiment will only
arise for narrow slits very close together). For regions $\Delta$
of small size, it may be necessary to consider coupling to an environment
to produce decoherence of histories. This will not be explored here.

\subsection{The Klein-Gordon Equation and the Probability for Surface Crossing}

We now consider the problem of assigning probabilities to histories
of the relativistic particle in flat space which
cross or do not cross the spacelike surface $x_0 = 0 $ in the spatial region
$\Sigma$. This is slightly different to the cases considered so far since we need
to show exactly how the results derived above for entering regions $\Delta$
may be used to derive probabilities for crossing surfaces. This example
also involves the effects of the signature of the metric. A decoherent histories
analysis of the Klein-Gordon system
was given in detail in Ref.\cite{HaTh1}, but it encountered difficulties
in definition of the class operators. A somewhat heuristic definition was used,
with satisfactory physical results, but the underlying origin of this definition
was not clear. Here, we show how this heuristic definition arises from
the more rigorous definitions used in this paper.

To define a class operator for crossing the spacelike surface $\Sigma$ ,
we regard the surface as
part of the boundary of a spacetime region $\Delta$. For simplicity we
take the three-dimensional region $\Sigma$ to be the interior of a $2$-sphere at $x^0 = 0 $. We then define $\Delta$ by requiring that it is a null cone in
$x^0 > 0 $ with $\Sigma $ its base.
(A two-dimensional version of this is depicted in Figure 7). The class operator is again given by Eq.(\ref{P15}),
with $ |\Psi \rangle $ a solution to the Klein-Gordon equation which may
contain positive and negative frequencies,
\beq
\Psi = \Psi^+ + \Psi^-
\eeq
The flux operator $ (\dot P)_{in}$ consists
of contributions from the two parts of the boundary of $\Delta$,
\beq
(\dot P)_{in} = (\dot P)_{in}^{\Sigma} + (\dot P)_{in}^{cone}
\eeq
The flux operator on $\Sigma$ is clearly just the positive frequency flux at this boundary
\beq
(\dot P)_{in}^{\Sigma} = \left( \theta ( \hat p_0 ) \hat p_0 \delta (\hat x^0 )
+  \delta (\hat x^0 ) \theta ( \hat p_0 ) \hat p_0 \right) \ f_{\Sigma} (\hat x^i )
\eeq
where $ f_{\Sigma} (x^i) $ is a window function localized in $\Sigma $. The remaining
part of the flux operator describes the incoming flux on the null cone.
Classically, since this surfaces is null, every incoming trajectory crossing
the cone will cross $\Sigma$. An analogous result is approximately
true in the quantum case, since propagation outside the lightcone is suppressed.
Hence, these flux operators are approximately equal to the negative frequency
flux at $\Sigma$,
\beq
(\dot P)_{in}^{cone} \approx
\left( \theta ( -\hat p_0 ) \hat p_0 \delta (\hat x^0 )
+  \delta (\hat x^0 ) \theta (- \hat p_0 ) \hat p_0 \right) \ f_{\Sigma} (\hat x^i )
\eeq
The class operator involves contributions from boths of these terms
and may be written
\beq
C_{\Sigma} | \Psi_{\l} \rangle = 2 \pi \delta ( H - \l )  \left(  |\hat p_0 |\delta (\hat x^0 )
+  \delta (\hat x^0 ) | \hat p_0 |\right) \ f_{\Sigma} (\hat x^i ) | \Psi_{\l} \rangle
\eeq
Noting that $ \langle x | \delta ( H ) | y \rangle $ is the same as the
the Klein-Gordon propagator $G^{(1)} = G^+ + G^- $ (where $G^{\pm}$ are the positive
and negative frequency Wightman functions \cite{HaTh1}),
we have, setting $\l = 0$,
\beq
\langle x | C_{\Sigma} | \Psi \rangle
= 2 \pi i \int_{\Sigma} d^3 y \left[ G^{(1)} (x,y) \overleftrightarrow {\partial_0 }\Psi^{+} (y)
- G^{(1)} (x,y) \overleftrightarrow {\partial_0} \Psi^{-} (y)
\right]
\eeq
This agrees with the definition in Ref.\cite{HaTh1}. It consists of the initial state attached
to $G^{(1)}$ with a ``sign-adjusted'' Klein-Gordon inner product to take care of the negative
frequencies. It was shown in Ref.\cite{HaTh1} that there is approximate decoherence between histories which
cross $x^0 = 0$ in $\Sigma$ or outside $\Sigma$, again because propagation outside
the lightcone is suppressed. The resulting probability for crossing is
\beq
p (\Sigma ) = 2 \pi i \int_{\Sigma} d^3 x \left[ \Psi^+ (x) \overleftrightarrow {\partial_0} \Psi^{+} (x)
- \Psi^- (x) \overleftrightarrow {\partial_0} \Psi^{-} (x) \right]
\eeq
This is the expected result and is positive.

Hence, the decoherent histories approach described in this paper yields the expected
results for the Klein-Gordon equation, improving on the heuristic analysis of Ref.\cite{HaTh1}.
Note that the method used above to relate probabilities for surface crossings to
probabilities for entering regions is specific to the dynamics of the relativistic
particle in flat space. More complicated systems will require a modified approach
adapted to their particular dynamics.

\section{WKB Regime}

The most important case in which to check the ideas developed above is in the WKB
regime. In the oscillatory regime, the solutions to the Wheeler-DeWitt equation
have the form
\beq
\Psi = R e^{ i S}
\label{W1}
\eeq
where $R$ and $S$ obey Eqs.(\ref{1.24}), (\ref{1.26}) as described earlier. More generally,
the wave function is a superposition of WKB wave functions but we consider first the case
of a single term. The heuristic interpretation of such states was described in
Section 1D. Our aim is to show that the decoherent histories analysis reproduces the
heuristic scheme. WKB states are locally plane wave states of the type
considered in the previous Section, so we will appeal to that analysis to understand the properties
WKB states.

We consider the action of the class operator for a single region
$\Delta$ on a WKB state. We first consider the class operator Eq.(\ref{P15})
acting on WKB state (regularized by making it an eigenstate of $H$ with eigenvalue $\l$)
\bea
\langle q | C_{\Delta} | \Psi_{\l} \rangle &=& 2 \pi \langle q | \delta ( H - \l ) (\dot P)_{in} | \Psi_{\l} \rangle
\nonumber \\
&=&
2 \pi i \int_{\Sigma_{in}} d^{n-1} q' \ \langle q | \delta ( H - \l ) | q' \rangle
 \overleftrightarrow {\partial_n} \Psi_{\l} (q')
\label{W2}
\eea
Here $\Sigma_{in}$ denotes the sections of the boundary where the flux is ingoing,
\beq
n \cdot \nabla S < 0
\eeq
where $n^a$ is the outward pointing normal. The key property of the WKB wave function $\Psi_{\l} (q')$
is that it has momentum $ p = \nabla S (q') $ at each point $q'$ on the boundary.
When the operator $\delta ( H - \l) $
is applied, from the representation Eq.(\ref{P8}), we see that it's effect is to evolve the state  $\Psi_{\l} (q')$
(restricted to $\Sigma_{in}$)
forwards and backwards in parameter time, and then integrates over all times. Semiclassically,
the evolution of the state $\Psi_{\l} (q')$ will be concentrated along the classical
trajectories defined by initial positions $q'$ in $\Sigma_{in}$ and momenta $ p = \nabla S (q') $,
with some spreading of the wave packet, but this will be small if $\Delta $ is reasonably
large. (This is analogous to the model of Section 5B).

We thus see the following: the wave function  $ \langle q | C_{\Delta} | \Psi_{\l},
\rangle $ in Eq.(\ref{W2})
is spatially localized around the tube of classical trajectories passing through
$ \Delta $ with momenta $ p  = \nabla S $ (depicted in Figure 2).
This may be approximately written in the alternative
form
\beq
\langle q^a | C_{\Delta} | \Psi_\l \rangle \approx \theta ( \tau_\Delta - \epsilon )
\  R e^{ i S}
\label{W3}
\eeq
where $ \epsilon > 0 $ is a small parameter to regularize the $\theta$-function at zero
argument.
Here, $\tau_\Delta (q) $ is the parameter time spent by the classical trajectory
$q_{cl} (t) $ (with initial value $q$ and momentum $ p = \nabla S (q) $) in the region $\Delta$ and may be written
\beq
\tau_{\Delta} (q) = \int_{-\infty}^\infty dt \ f_{\Delta} ( q_{cl} (t) )
\eeq
This has the property that
\beq
\nabla S \cdot \nabla \tau_{\Delta} = 0
\eeq
since $\nabla S \cdot \nabla $ simply translates along the classical trajectories.
It follows from Eq.(\ref{1.26}) that Eq.(\ref{W3}) is in fact a WKB solution
to the Wheeler-DeWitt equation, since the $\theta$-function essentially modifies the prefactor
$ R $ but in a way that it still satisfies Eq.(\ref{1.26}). (Of course, this is expected because $C_{\Delta}$
commutes with $H$).

Eq.(\ref{W3}) is a very useful result and allows us to check for decoherence very easily.
The action of the class operator for not entering $\Delta $ is clearly
\beq
\langle q^a | \bar C_{\Delta} | \Psi_\l \rangle \approx \theta ( \epsilon - \tau_\Delta )
\  R e^{ i S}
\label{W6}
\eeq
which is a WKB state localized on the set of trajectories not entering $\Delta$.
It immediately follows that
\beq
\langle \Psi_{\l'} | \bar C_{\Delta} C_{\Delta} | \Psi_{\l} \rangle \approx 0
\eeq
since the two states Eqs.(\ref{W3}), (\ref{W6}) are localized about complementary regions. There is
therefore approximate decoherence of histories for a single WKB packet and for histories entering or not entering
a single region $\Delta$, as long as $\Delta $ is sufficiently large.

The key reason for the decoherence with a single WKB packet
is related to the approximate determinism of the WKB wave functions: fixing values of position
to lie on $\Sigma_{in}$ also fixes the momenta, since $ p = \nabla S $, so that
Eq.(\ref{W2}) is concentrated along a tube of classical trajectories.

More generally, the initial state will be a superposition of WKB wave packets,
\beq
\Psi = \sum_{k} R_k e^{ i S_k }
\eeq
The component states in this sum are typically approximately orthogonal to each other as long as the
phases $S_k$ are sufficiently different. (This will depend on the detailed dynamics
of the model). Following the analogous example in the simple models of Section 5B, we would
expect that the class operators will not disturb the approximate orthogonality of these
states as long as the region $ \Delta $ is sufficiently large -- again we expect
the cross terms in the decoherence functional to average to zero because of oscillations,
as in Eq.(\ref{5.30}). Therefore, superpositions may in practice be treated as mixtures
at sufficiently coarse-grained scales. Note that this statement also applies the
special state,
\beq
\Psi = R \left( e^{ i S} + e^{ - i S} \right)
\eeq
which arises from the Hartle-Hawking ``no boundary'' proposal \cite{HarHaw}. That is, the
interference between the two terms may be neglected.

Given decoherence, the probabilities are given by the general expression Eq.(\ref{P17}).
It then follows that the probabilities for entering $\Delta$ coincide with Eq.(\ref{1.29}),
the sought-after result.

Now consider the case of probabilities for histories entering two regions, as described
by the (modified) class operator Eq.(\ref{P36}). Since the two-region class operator is a sum of
products of one-region class operators, its effect on the WKB wave functions is easy
to see. The action of a single class operator gives Eq.(\ref{W3}). But since this is still
a wave function of the WKB type, the action of a second class operator yields
\beq
\langle q^a | C_{\Delta_1 \Delta_2} | \Psi_\l \rangle \approx
\theta ( \tau_{\Delta_1} - \epsilon ) \theta ( \tau_{\Delta_2} - \epsilon )
\  R e^{ i S}
\label{W3b}
\eeq
That is, it is a WKB wave function but restricted in such as way that it's flux passes
through both regions. See Figure 8. (It is easy to see that in this case the action
of the unmodified class operator Eq.(\ref{P34}) is the same.)
It is again easy to see that there is decoherence
of histories and the probability is given by an expression of the form Eq.(\ref{1.29}), but
with the integral over the subset of incoming flux at $\Delta_1$ which goes
on to intersect $\Delta_2$.

From the above observations one can also see why the exponential WKB wave functions will {\it not} lead to
decoherence of histories. The exponential wave functions have the form
\beq
\Psi = R \ e^{-I}
\eeq
where $I$ is real. The key difference between states of this type and the oscillatory type
Eq.(\ref{W1}) is that they do not have a correlation between positions and momenta \cite{Halcor}.
One would
therefore expect the evolution of the state Eq.(\ref{W2}) to be spread all over the configuration space
and not concentrated around a particular region. That is, these states do not have the approximate
determinism of the oscillatory states. The states $ C_{\Delta} | \Psi \rangle$
and $ \bar C_{\Delta} | \Psi \rangle $ would then not be approximately orthogonal so there will
be no decoherence of histories.

The decoherence of histories described here has arisen because of the approximate determinism
of the oscillatory WKB states, together with the approximate orthogonality properties that
arise when the regions $\Delta$ are sufficiently large. At finer grained scales, decoherence
of histories may only be possible in more complicated models in which there is an environment
of some sort. Models along these lines, in more basic approaches to quantum cosmology,
have been considered previously \cite{decoinqc}, and one might expect that they may
be adapted to the decoherent histories approach to quantum cosmology. (See also Ref.\cite{HaTh2}).

In summary, we have derived from the decoherent histories approach
the probabilities normally used in the heuristic WKB interpretation.
To address these issues in more detail will require more specific quantum cosmological
models. This will be considered elsewhere.

\section{Discussion and Further Issues}

We have presented a properly defined quantization procedure
for quantum cosmology using the decoherent histories approach to quantum
theory and derived from this the frequently used but heuristic WKB interpretation,
involving fluxes of the WKB wave function.

The key idea was to use a complex potential to define the class operators
for not entering a region of configuration space. This method is adequately justified
by its successful use in the arrival time problem in non-relativistic quantum
theory. We showed that the class operators defined in this way have all the
desired properties -- they have the correct classical limit, are compatible
with the constraint equation, and do not have difficulties with the Zeno effect.
In a semiclassical approximation, they have an appealing form in terms
of intersection number operators. They give sensible results in simple
models and there is approximate decoherence of histories for certain types
of initial state at sufficiently coarse grained scales.

Future papers will address the more detailed application of this approach
to specific models and will also undertake a comparison of the decoherent histories approach
described here to other approaches \cite{Rov}.

%SHO models. Comparison with earlier results of Hartle.

\section{Acknowledgements}

I am very grateful to Jim Hartle, Don Marolf and James Yearsley for useful conversations.
I would also like to thank James Yearsley for his help in preparing the figures.

\appendix

\section{Operators and Inner Products for the Klein-Gordon Equation}

In this appendix we outline the relationship between the induced inner
product and Klein-Gordon inner product. This has been described elsewhere
(see, for example, Ref.\cite{HaMa}),
but the derivation given here is different and will help give some insight
into the form of some of the results of Section 4.

We use $x^{\mu} = (x^0, \x) $ to denote spatial coordinates and take the
signature of spacetime to be $ (- + + + )$.
The Hamiltonian for the Klein-Gordon system is
\beq
H = - p_0^2 + \p^2 + m^2
\eeq
The induced inner product begins by looking at eigenfunctions of $H$
with eigenvalue $\l$, which we write as
\bea
\langle x | \l \p s \rangle &=& \Psi_{\l \p s} (x)
\nonumber \\
&=& N e^{ i s p \cdot x}
\eea
where $p_0 = \sqrt{ \p^2 + m^2 - \l} $ and $s= \pm 1$ labels the solutions as positive or negative
frequency.
We work with three different inner products. The first is the
standard Schr\"odinger inner product,
\beq
\langle \l \p s | \l' \p' s' \rangle =
\int d^4 x \ \Psi^*_{ \l \p s} (x) \Psi_{\l' \p' s'} (x)
\label{A3}
\eeq
and the eigenfunctions are normalized with this inner product according to
\beq
\langle \l \p s | \l' \p' s' \rangle =
\delta (\l - \l')\ \delta ( \p - \p') \ \delta_{s s'}
\label{A4}
\eeq
The goal is to relate this to the Klein-Gordon inner product Eq.(\ref{1.6}).
One way to do this is to integrate out $x^0$ in Eq.(\ref{A3}), but here
we proceed differently.

We introduce the projection operator $ P = \theta ( \hat x^0)$ onto $x^0 > 0 $
and consider the current operator
\beq
\hat J = \int_{-\infty}^{\infty} dt \ \dot P(t)
\eeq
On the one hand, we can carry out the integral with the result
\bea
\hat J &=& \left[ \theta ( \hat x^0 + 2 \hat p^0 t ) \right]^{t=\infty}_{t=- \infty}
\nonumber \\
&=& \epsilon ( \hat p^0 )
\eea
where $ \epsilon ( \hat p^0 ) $ is the signum function operator. On the other hand, we can differentiate
$P$ with the result
\bea
\dot P (t)  &=& i [ H, \theta ( \hat x^0  )]
\nonumber \\
&=& \hat p^0 \delta (\hat x^0 )
+ \delta ( \hat x^0  ) \hat p^0
\label{A7a}
\eea
so that
\beq
\hat J =  \int_{-\infty}^{\infty} dt \ e^{ i H t } \left( \hat p^0 \delta (\hat x^0)
+ \delta ( \hat x^0 ) \hat p^0 \right) e^{ - i H t }
\label{A7}
\eeq
We now insert the identity
\beq
\epsilon (\hat p^0 ) \hat J = 1
\eeq
with $\hat J$ given by Eq.(\ref{A7}) in the inner product Eq.(\ref{A4}), with the result
\bea
\langle \l \p s | \l' \p' s' \rangle &=&
\langle \l \p s |\ \epsilon (\hat p^0 ) \hat J \ | \l' \p' s' \rangle
\nonumber \\
&=& 2 \pi \delta (\l - \l') \ s \
\langle \l \p s | \left( \hat p^0 \delta (\hat x^0)
+ \delta ( \hat x^0 ) \hat p^0 \right) |\l \p' s' \rangle
\label{A9}
\eea
We now identify the inner product expression on the right-hand side as the
Klein-Gordon inner product between states with the same $\l$,
\bea
( \Psi_{\l \p s}, \Psi_{\l \p' s'} )_{KG} &\equiv & \langle \l \p s | \left( \hat p^0 \delta (\hat x^0)
+ \delta ( \hat x^0 ) \hat p^0 \right) |\l \p' s' \rangle
\nonumber \\
&=& i \int_{x^0=0}  d^3x \   \Psi^*_{\l \p s} (x)  \overleftrightarrow {\partial_0} \Psi_{\l \p' s'} (x)
\label{A10}
\eea
(where $ \hat p^0 = - \hat p_0 = i \partial_0 $).
Therefore the relationship between the Schr\"odinger inner product and Klein-Gordon inner product
is
\beq
\langle \l \p s | \l' \p' s' \rangle = 2 \pi \delta (\l - \l') \ s \
( \Psi_{\l \p s}, \Psi_{\l \p' s'} )_{KG}
\eeq
Finally, the induced (or physical) inner product on eigenstates with the same $\l$,
is defined as outlined in Section 1,
essentially by factoring out the $\delta$-function
\beq
\langle \l \p s | \l' \p' s' \rangle = \delta ( \l - \l') \langle \l \p s | \l \p' s' \rangle_{phys}
\eeq
so we deduce that
\beq
\langle \l \p s | \l \p' s' \rangle_{phys} = 2 \pi s \ ( \Psi_{\l \p s}, \Psi_{\l \p' s'} )_{KG}
\eeq
and we may set $\l=0$ to obtain an inner product on solutions to the Klein-Gordon equation.
The Klein-Gordon inner product is of course positive on positive frequency ($s=1$) solutions
and negative on negative frequency ($s=-1$) solutions, but the form of the above result, with the overall
factor of $s$, ensures that the induced inner product is always positive.

Note that some caution is required in relation to Eqs.(\ref{A7a}) and (\ref{A9}), since
Eq.(\ref{A7a}) implies that
\bea
\langle \l \p s | \left( \hat p^0 \delta (\hat x^0)
+ \delta ( \hat x^0 ) \hat p^0 \right) |\l' \p' s' \rangle
&=& i \langle \l \p s | [ H, \theta (\hat x^0)] |\l' \p' s' \rangle
\nonumber \\
&=& i ( \l - \l') \langle \l \p s |  \theta (\hat x^0)|\l' \p' s' \rangle
\label{A15}
\eea
This appears to be zero when $\l = \l'$ suggesting that Eq.(\ref{A9}) is zero.
This is not in fact the case since the inner product
expression on the right-hand side of Eq.(\ref{A15}) is actually divergent as $ \l \rightarrow \l'$.

\bibliography{apssamp}

\begin{thebibliography}{10}

\bibitem{QC1} J.J.Halliwell, in
Quantum Cosmology and Baby Universes, edited by S.Coleman, J.B.Hartle, T.Piran
and S.Weinberg (World Scientific, Singapore, 1991) also available as the
e-print gr-qc/0909.2566. These introductory lectures on quantum cosmology
contain a detailed account of the WKB interpretation.

\bibitem{QC} C.Kiefer and B.Sandhoefer, gr-qc/0804.0672; D.L.Wiltshire, gr-qc/0101003.

\bibitem{QC2} J.B.Hartle, S.W.Hawking and T.Hertog, Phys.Rev.D77, 123537 (2008);
Phys.Rev.Lett.100, 201301 (2008).


\bibitem{DeW} B.DeWitt, in {\it Gravitation: An Introduction to
Current Research}, edited by L.Witten (John WIley and Sons, New
York, 1962).
% The quantization of geometry.

\bibitem{Rov} C.Rovelli, {Class.Quant.Grav.} {8}, 297 (1991);
% What is observable in classical and quantum gravity
{8}, 317 (1991);
% Quantum references systems.
Phys.Rev. 42, 2638 (1990);
% Quantum mechanics without time: A model.
Phys.Rev. 43, 442 (1991);
% Time in quantum gravity: An hypothesis.
D.Marolf, Class.Quant.Grav. {12}, 1199 (1995);
% Quantum observables and recollapsing dynamics
{12}, 1441 (1995);
% Observables and a Hilbert Space for Bianchi IX
P.Hajicek, J.Math.Phys. { 36}, 4612 (1995);
M.Montesinos,C.Rovelli and T.Thiemann, Phys.Rev. {D60}, 044009 (1999);
M.Montesinos, Gen.Relativ.Gravit. {33}, 1 (2001);
R.Gambini and R.A.Porto, Phys.Rev. { D63}, 105014 (2001).
% Relational time in generally covariant quantum systems: four models




\bibitem{Time} K.Kuchar, gr-qc/9304012 and in {\it Proceedings of the 4th Canadian Conference on
General Relativity and Relativistic Astrophysics} edited by
G.Kunstatter, D.Vincent and J.Williams (World Scientific,
Singapore, 1992); C.Isham, gr-qc/9210011.


\bibitem{FRW} S.W.Hawking, Nucl.Phys. B239, 257 (1984).
%The quantum state of the universe.

\bibitem{Har3} J.B.Hartle, in {\it Proceedings of the 1992 Les Houches
School, Gravity and its Quantizations}, edited by B.Julia and
J.Zinn-Justin (Elsevier Science B.V., 1995)
% Spacetime quantum mechanics and the quantum mechanics of
% spacetime

\bibitem{HaMa} J.B.Hartle and D.Marolf,  {Phys.Rev.} {D56}, 6247 (1997).
% Comparing Formulations of Generalized Quantum Mechanics for
% Reparametrization-Invariant Systems

\bibitem{HaTh1} J.J.Halliwell and J.Thorwart, {Phys.Rev.} {D64}, 124018 (2001).
% Decoherent histories analysis of the relativistic particle

\bibitem{HaTh2} J.J.Halliwell and J.Thorwart, Phys.Rev. {D65}, 104009 (2002)
% Life in an Energy Eigenstate: Decoherent Histories Analysis of a
% Model Timeless Universe

\bibitem{HaWa} J.J.Halliwell and P.Wallden,
Phys. Rev. D73, 024011 (2006)


\bibitem{CrHa} D.Craig and J.B.Hartle,
Phys. Rev. {D69}, 123525 (2004).
% Generalized quantum theory of recollapsing homogeneous cosmologies



\bibitem{Whe} J.Whelan, {Phys.Rev.} {D50}, 6344
(1994).
% Spacetime alternatives in relativistic particle motion




\bibitem{AnSa} C.Anastopoulos and K.Savvidou,
Class.Quant.Grav. {17}, 2463 (2000);
%Histories quantisation of
%parameterised systems: I. Development of a general algorithm
{22}, 1841 (2005).
%Minisuperspace Models in Histories Theory




\bibitem{Rie} A.Ashtekar, J.Lewandowski, D.Marolf, J.Mourao and
T.Thiemann, {J.Math.Phys.} {36}, 6456 (1995);
% Quantization of diffeomorphism invariant theorie of connections with
% local degrees of freedom.
A.Higuchi, {Class.Quant.Grav.}  8, 1983 (1991).
% Quantum linearization instabilities of de Sitter spacetime 2.
D.Giulini and D.Marolf, {Class.Quant.Grav.} {16}, 2489 (1999);
% A Uniqueness Theorem for Constraint Quantization
{Class.Quant.Grav.} {16}, 2479 (1999).
% On the Generality of Refined Algebraic Quantization
F.Embacher, {Hadronic J.} {21}, 337 (1998);
% Handwaving refined algebraic quantization.
N.Landsmann, { J.Geom.Phys.} {15}, 285 (1995);
% Rieffel induction as generalized quantum Marsden-Weinstein
% reduction
D.Marolf, gr-qc/0011112.
% Group Averaging and Refined Algebraic Quantization: Where are we now?


\bibitem{GH1} M.Gell-Mann and J.B.Hartle, in {\it Complexity, Entropy
and the Physics of Information, SFI Studies in the Sciences of
Complexity}, Vol. VIII, W. Zurek (ed.) (Addison Wesley, Reading,
1990); and in {\it Proceedings of the Third International
Symposium on the Foundations of Quantum Mechanics in the Light of
New Technology}, S. Kobayashi, H. Ezawa, Y. Murayama and S. Nomura
(eds.) (Physical Society of Japan, Tokyo, 1990).
%{\it Quantum Mechanics in the Light of Quantum Cosmology.}

\bibitem{GH2} M.Gell-Mann and J.B.Hartle, Phys.Rev. {D47},
3345 (1993).
%{\it Classical Equations for Quantum Systems}.

\bibitem{Gri} R.B.Griffiths, J.Stat.Phys. {36}, 219 (1984);
Phys.Rev.Lett. {70}, 2201 (1993); Phys.Rev. {A54}, 2759
(1996); {A57}, 1604 (1998).

\bibitem{Omn1} R. Omn\`es, J.Stat.Phys. {53}, 893 (1988).
{53}, 933 (1988); {53}, 957 (1988); {57}, 357 (1989);
Ann.Phys. { 201}, 354 (1990); Rev.Mod.Phys. 64, 339
(1992).

\bibitem{Omn2} R.Omn\`es, J. Math. Phys. {38}, 697 (1997)
% Quantum-classical correspondence using projection operators


\bibitem{Hal1} J.J.Halliwell, in {\it Fundamental Problems in Quantum
Theory},  edited by D.Greenberger and A.Zeilinger, Annals of the
New York Academy of Sciences, {775}, 726 (1994).

\bibitem{Hal2} J.J.Halliwell, e-print quant-ph/0301117, to appear
in Proceedings of the Conference, Decoherence, Information,
Complexity, Entropy (DICE), Piombino, Italy, September, 2002
(edited by T.Elze).

\bibitem{Halrec} J.J.Halliwell, Phys.Rev. D60, 105031 (1999).


\bibitem{Ish} C. Isham, {J.Math.Phys.} {23}, 2157
(1994);
%{\it Quantum Logic and the Histories Approach to Quantum Theory.}
C. Isham and N. Linden, {J.Math.Phys.} {35}, 5452 (1994);
{36}, 5392 (1995).

\bibitem{HalKK} J.J.Halliwell Nucl.Phys. B266, 228 (1986);
B286, 729 (1987).


\bibitem{Teit} J.J.Halliwell, Phys.Rev. D38, 2468 (1988);
J.J.Halliwell and J.B.Hartle, Phys.Rev.
{ D41}, 1815 (1990); { D43}, 1170 (1991); C.Teitelboim, Phys.Rev. D25, 3159 (1982);
D28, 297 (1983).


\bibitem{Halcor} J.J.Halliwell,
Phys. Rev. D36, 3626 (1987).


\bibitem{timeinqm} J.G.Muga, R.Sala Mayato and I.L.Egusquiza (eds),
Time in Quantum Mechanics (Springer, Berlin, 2002); J.G.Muga and
C.R.Leavens, Phys.Rep.338, 353 (2000).


\bibitem{complex} J.G.Muga, S.Brouard, D.Macias, Ann. Phys. (NY) 240, 351 (1995);
Ph.Blanchard and A.Jadczyk, Helv. Phys. Acta 69, 613 (1996);
J.P.Palao, J.G.Muga, S.Brouard, A.Jadczyk, Phys. Lett. A233, 227
(1997); A.Ruschhaupt, J.Phys. A35, 10429 (2002); J.G. Muga, J.P.
Palao, B. Navarro, I.L. Egusquiza, Phys. Rep. 395, 357 (2004);
J.J.Halliwell, Prog.Theor.Phys. 102, 707 (1999).



\bibitem{HaYe1} J.J.Halliwell and J.M.Yearsley, Phys. Rev. A79, 062101 (2009).


\bibitem{HaYe2} J.J.Halliwell and Y.Yearsley, quant-ph/0903.1958.


\bibitem{YaT} N.Yamada and S.Takagi, {Prog.Theor.Phys.}
{85}, 985 (1991); {86}, 599 (1991); {87}, 77 (1992);
N. Yamada, { Sci. Rep. T\^ohoku Uni., Series 8}, {12}, 177
(1992); {Phys.Rev.} {A54}, 182 (1996).

\bibitem{HaZa} J.J.Halliwell and E.Zafiris, {Phys.Rev.} {D57},
3351-3364 (1998).

\bibitem{Har4} J.B.Hartle, {Phys.Rev.} {D44}, 3173 (1991).
%{\it Spacetime Coarse-Grainings in Non-Relativistic Quantum Mechanics.}
%R.J.Micanek and J.B.Hartle, {Phys.Rev.} {A54}, 3795 (1996).
% Nearly instantaneous alternatives in quantum mechanics.





\bibitem{Ech} J.Echanobe, A. del Campo and J.G.Muga, Phys. Rev. A77, 032112
(2008).

\bibitem{Ye1} J.Yearsley, in preparation.



\bibitem{PDX} A.Auerbach and S.Kivelson, {Nucl.Phys.}
B257, 799 (1985); P. van Baal, in {\it Lectures on Path
Integration}, edited by H.A. Cerdeira et al. (World Scientific,
Singapore, 1993) [also available from the website
http://www.lorentz.leidenuniv.nl/vanbaal/HOME/publ.html];

\bibitem{HaOr} J.J.Halliwell and M.E.Ortiz, {Phys.Rev.}
D48, 748 (1993).


\bibitem{Hal3} J.J.Halliwell, Phys.Lett. A207, 23 (1995).


\bibitem{Haltraj} J.J.Halliwell, Phys.Rev. D64, 044008 (2001).
%``Trajectories for the Wave Function of the Universe from a
%Simple Detector Model'', % gr-qc/0008046.




\bibitem{Sch} L.S.Schulman, Techniques and Applications of Path Integration
(Wiley and Sons, New York, 1981), chapter 7.


\bibitem{Mott} N.F.Mott, {Proc.Roy.Soc} {A124}, 375 (1929),
% The wave mechanics of alpha-ray tracks.
reprinted in {\it Quantum Theory and Measurement}, edited by
J.Wheeler and W.Zurek (Princeton University Press, Princeton, New
Jersey, 1983). For further discussions of the Mott calculation see
also, J.S.Bell, {\it Speakable and Unspeakable in Quantum
Mechanics} (Cambridge University Press, Cambridge, 1987);
A.A.Broyles, Phys.Rev. {A48}, 1055 (1993); M.Castagnino and
R.Laura,  Int.J.Theor.Phys. {39}, 1737 (2000); J.Barbour,
Class.Quant.Grav. {11}, 2853 (1994); {11}, 2875 (1994).

\bibitem{HarHaw} J.B.Hartle and S.W.Hawking, Phys. Rev. D28, 2960
(1983).

\bibitem{decoinqc} J.J.Halliwell, Phys.Rev. D39, 2912 (1989);
{C.Kiefer}, Class.Quant.Grav. 4, 1369 (1987);
{T.Padmanabhan}, Phys.Rev. D39, 2924 (1989).






%\bibitem{Har10} J.Hartle, Class. Quantum Grav. { 13}, 361
%(1996).





%\bibitem{KaNa} Y.Kazama and R.Nakayama, {Phys.Rev.} { 32}, 2500 (1985).
% Wave packet in quantum cosmology.

%\bibitem{Kie} C.Kiefer, {Phys.Rev.} { D38}, 1761 (1988).
% Wave packets in minisuperspace.

%\bibitem{Kla} J.Klauder, Ann. Phys. (NY) { 254}, 419 (1997),
%(quant-ph/9604033) ; Nucl.Phys. { B547},397, 1999,
%(hep-th/9901010); hep-th/0003297.





\end{thebibliography}

\vfil \eject

\epsfxsize=5in\epsfbox{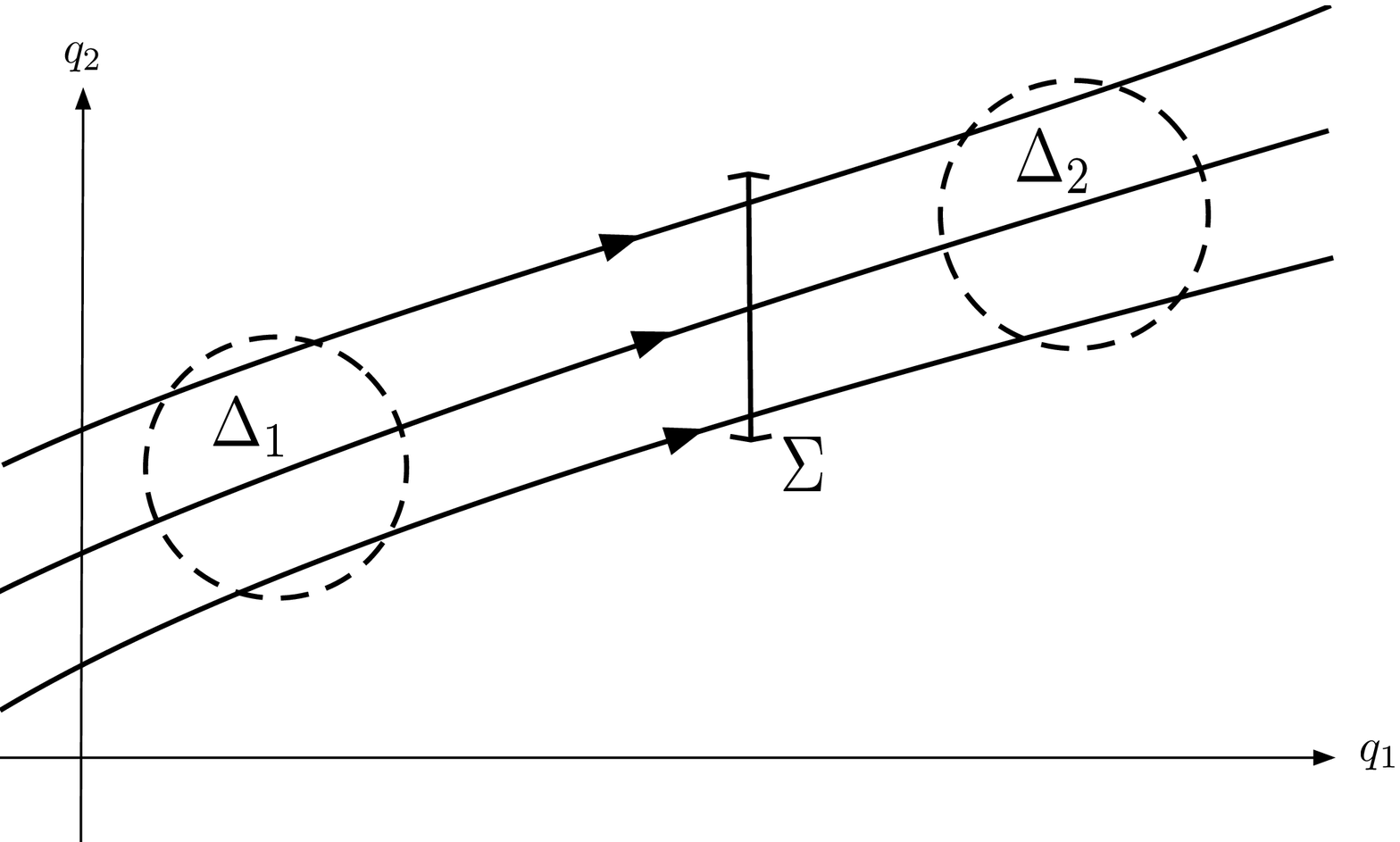}

\noindent{\bf Figure 1.} Given a solution $ \Psi $ to the Wheeler-DeWitt equation,
what is the probability that the system enters a series of regions $\Delta_1, \Delta_2 \cdots$
in configuration space, or crosses a surface $\Sigma$, at any stage in the system's
entire history?

\bigskip
\bigskip

\epsfxsize=5in\epsfbox{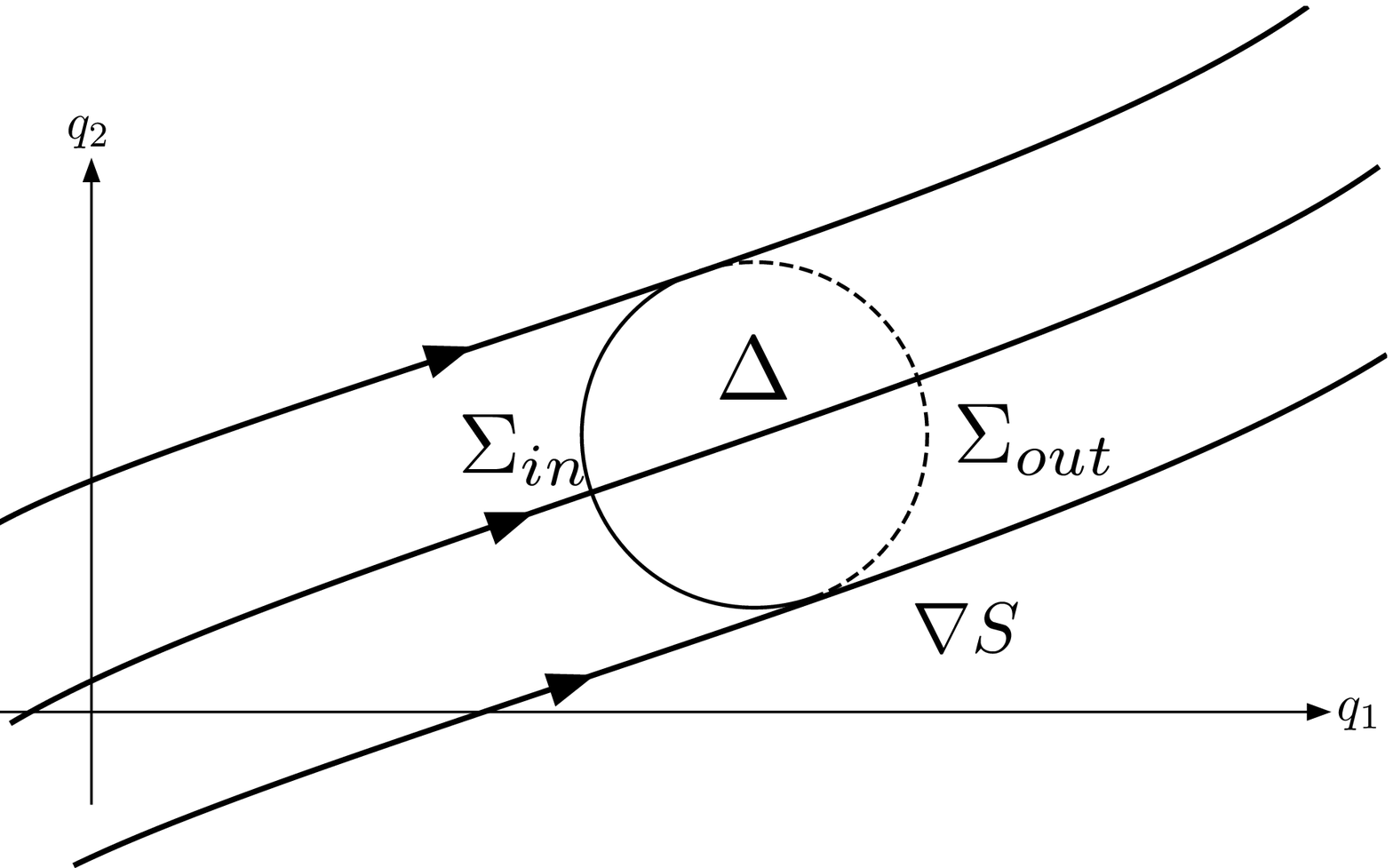}

\noindent{\bf Figure 2.} A WKB wave function with phase $S$ corresponds to a set
of classical trajectories with tangent vector $\nabla S$.
The probability for entering a region $\Delta$ is the amount flux of the wave function
intersecting $\Delta$. It may be expressed either in terms of the ingoing
flux across $\Sigma_{in}$ or equally, in terms of the outgoing flux
across $\Sigma_{out}$.

\vfil \eject

\epsfxsize=5in\epsfbox{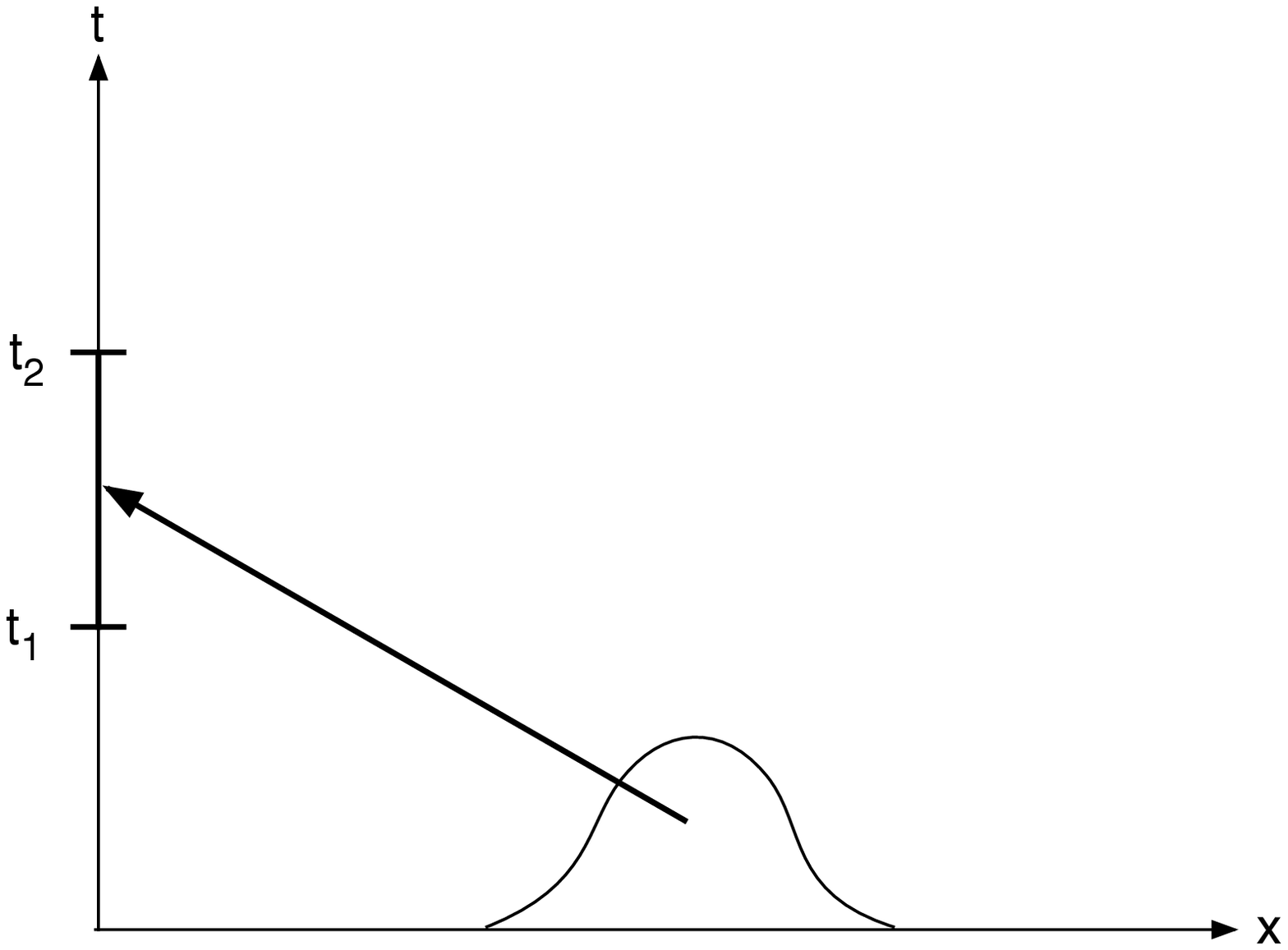}

\noindent{\bf Figure 3.} The quantum arrival time problem in non-relativistic quantum
mechanics. Given an initial state
localized entirely in $x>0$ and consisting entirely of negative momenta, what is
the probability that the particle crosses the origin during the time interval $[t_1,t_2]$?

\bigskip
\bigskip
\epsfxsize=5in\epsfbox{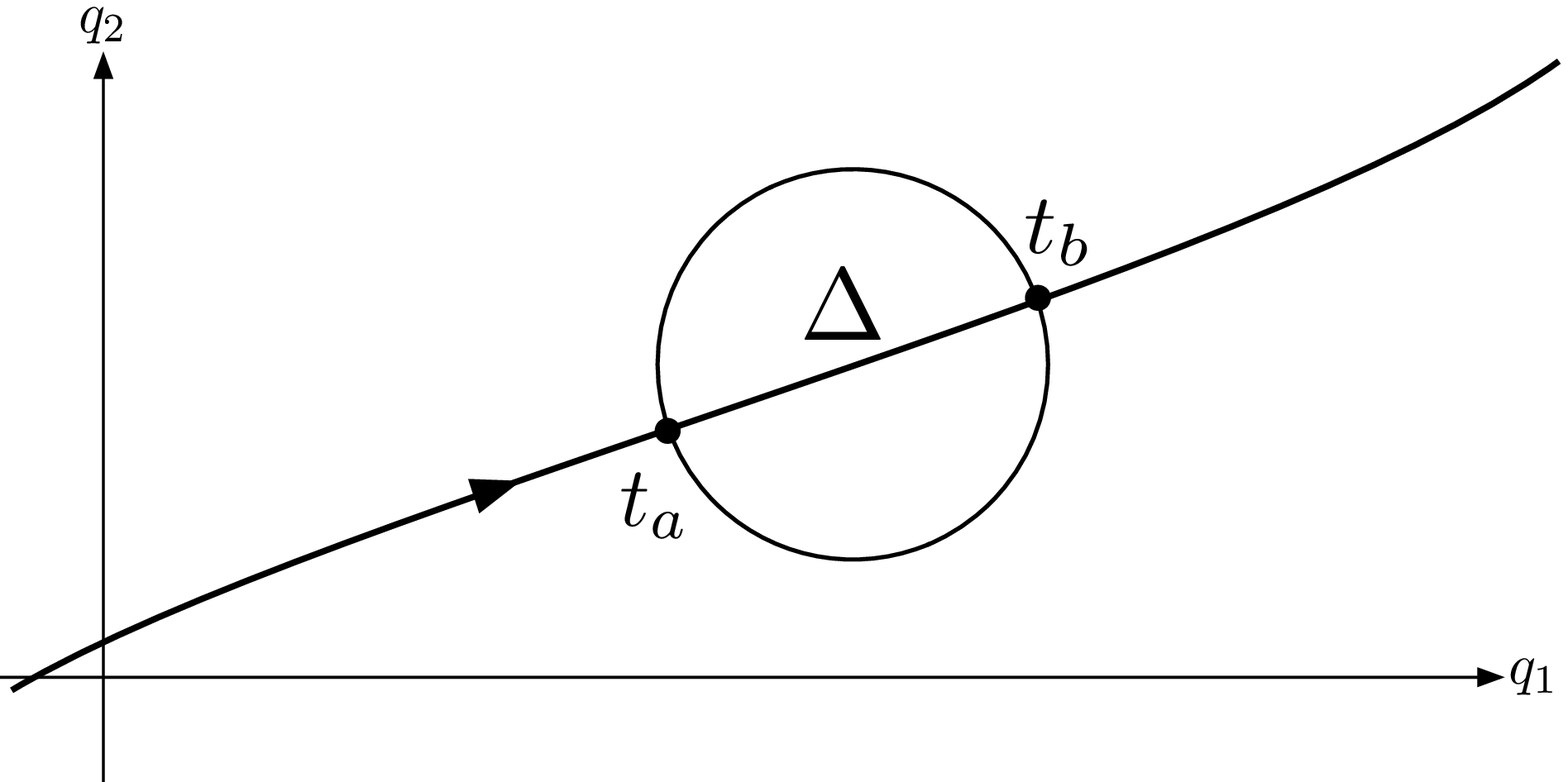}

\noindent{\bf Figure 4.} A classical trajectory $q^a(t)$ intersecting $\Delta$
enters at parameter time $t_a$ and leaves at parameter time $t_b$.

\vfil \eject

\epsfxsize=5in\epsfbox{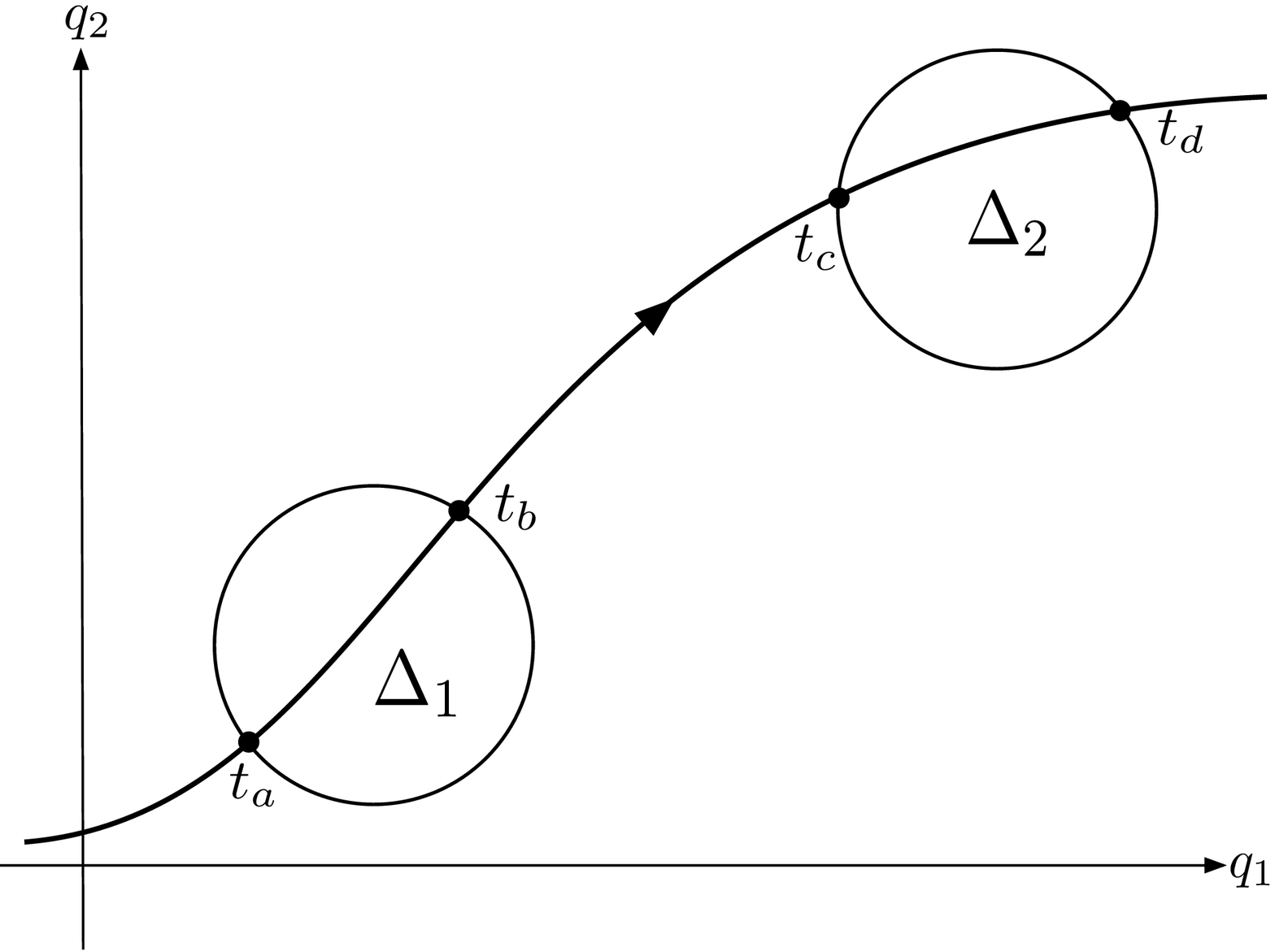}

\noindent{\bf Figure 5.} A classical trajectory $q^a(t)$ intersecting $\Delta_1 $ and
$\Delta_2$ enters $\Delta_1$ at $t_a$, leaves at $t_b$, enters $ \Delta_2 $
at $t_c$ and leaves at $t_d$.

\bigskip
\bigskip

\epsfxsize=5in\epsfbox{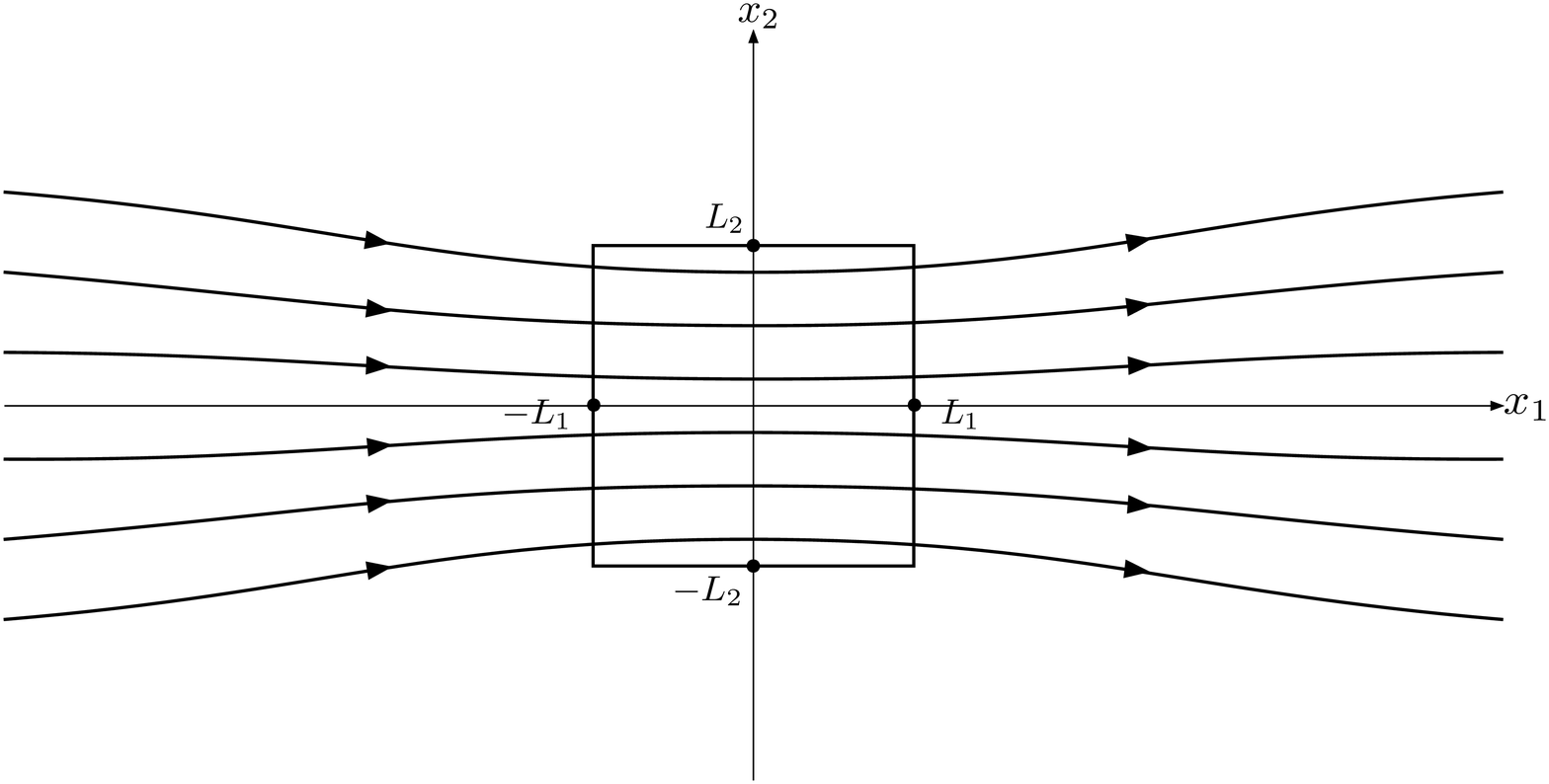}

\noindent{\bf Figure 6.} The class operator $C_{\Delta}$ acting on the initial
state $ | \Psi \rangle $ produces a state consisting of a plane wave $ \exp ( i p_1 x_1) $
localized along a tube of width
$ 2 L_2$ centred around the $x_1$ axis, with spreading through a small
angle as $x_1 \rightarrow \pm \infty $.

\vfil \eject

\epsfxsize=5in\epsfbox{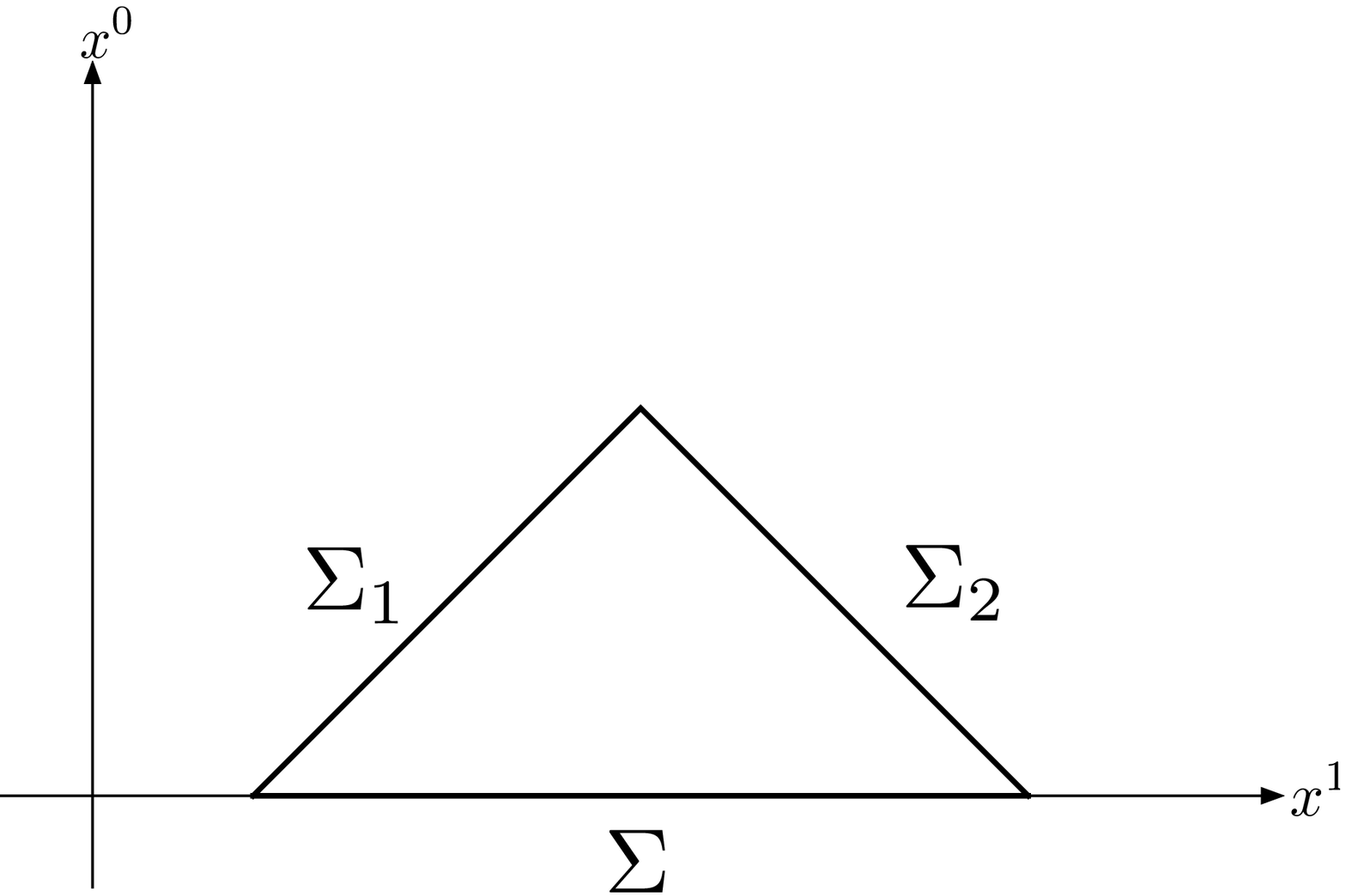}

\noindent{\bf Figure 7.} The class operator for a relativistic particle
to cross a surface $\Sigma$ may be calculated by regarding $\Sigma$
as part of the boundary of the spacetime region $\Delta$, a section of null
cone whose base is $\Sigma$. In two dimensions, depicted here, it is a triangle with base
$\Sigma$ whose other two
sides are the null surfaces $\Sigma_1, \Sigma_2$.

\bigskip
\bigskip

\epsfxsize=5in\epsfbox{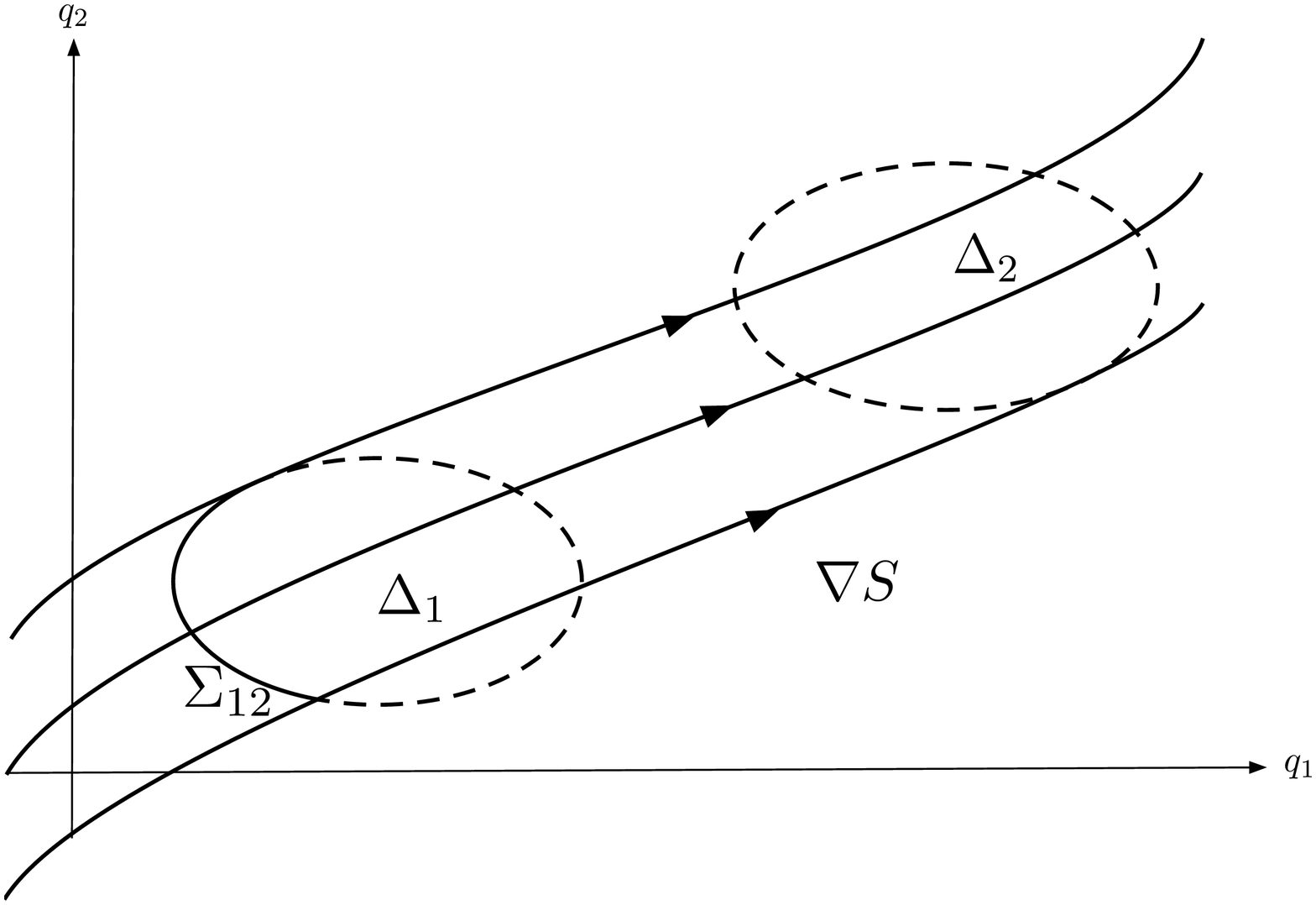}

\noindent{\bf Figure 8.}  The class operator $C_{\Delta_1 \Delta_2}$ for
two regions operating on a WKB wave function produces another WKB state
localized around the flux passing through both regions. This is the same
as the flux entering $\Delta_1$ across the surface $\Sigma_{12}$.

\end{document}